\documentclass[pra,twocolumn,superscriptaddress]{revtex4}

\usepackage{color}
\usepackage{amsfonts}
\usepackage{amsmath}
\usepackage{amssymb}
\usepackage{graphicx}

\newcommand{\be}{\begin{equation}}
\newcommand{\ee}{\end{equation}}
\newcommand{\bea}{\begin{eqnarray}}
\newcommand{\eea}{\end{eqnarray}}

\begin{document}

\title{Emergent chirality in multi-lead Luttinger-liquid junctions out of equilibrium}
\author{D. N. Aristov}
\affiliation{NRC ``Kurchatov Institute", Petersburg Nuclear Physics Institute, Gatchina 188300, Russia}
\affiliation{Institute for Nanotechnology, Karlsruhe Institute of Technology, 76021 Karlsruhe, Germany}
\affiliation{St.Petersburg State University, 7/9 Universitetskaya nab., 199034 St.\,Petersburg, Russia}
\author{I. V. Gornyi}
\affiliation{Institute for Nanotechnology, Karlsruhe Institute of Technology, 76021 Karlsruhe, Germany}
\affiliation{A.F. Ioffe Physico-Technical Institute, 194021 St.\,Petersburg, Russia}
\author{D. G. Polyakov}
\affiliation{Institute for Nanotechnology, Karlsruhe Institute of Technology, 76021 Karlsruhe, Germany}
\author{P. W\"olfle}
\affiliation{Institute for Nanotechnology, Karlsruhe Institute of Technology, 76021 Karlsruhe, Germany}
\affiliation{Institute for Condensed Matter Theory, Karlsruhe Institute of Technology, 76128 Karlsruhe, Germany}

\begin{abstract}
We study charge transport through $N$-lead junctions ($N\geq 3$) of spinless Luttinger liquid wires with bias voltages applied to Fermi-liquid reservoirs. In particular, we consider a Y junction, which is a setup characteristic of the tunneling experiment. In this setup, the strength of electron-electron interactions in one of the arms (``tunneling tip") is different from that in the other two arms (which form together the ``main wire"). For a generic single-particle $S$ matrix of the junction, we find that the bias voltage $V$ applied---even symmetrically---to the main wire generates a current proportional to $|V|$ in the tip wire. We identify two mechanisms of this nonequilibrium-induced ``emergent chirality" in a setup characterized by the time-reversal and parity symmetric Hamiltonian of the junction. These are: (i) the emergence of an effective magnetic flux, which breaks time-reversal symmetry, and (ii) the emergence of parity-breaking asymmetry of the setup, both proportional to the interaction strength and the sign of the voltage. The current in the tip wire generated by mechanism (i) is reminiscent of the Hall current in the linear response of a system the Hamiltonian of which breaks time-reversal symmetry; however, in the absence of any magnetic field or a local magnetic moment. Similarly, mechanism (ii) can be thought of as an emergent ``photogalvanic effect"; however, in the presence of inversion symmetry within the main wire. The nonequilibrium chirality implies a rectification of the current in the tip when the main wire is biased by {\it ac} voltage.
\end{abstract}

\maketitle

\section{Introduction}
\label{s1}

The transport properties of electric circuits built out of single-channel interconnected quantum wires are strongly affected by the peculiar charge screening at the junctions, which leads to critical behavior of the electric current as a function of the wire length, temperature, or bias voltage
\cite{Kane1992,Furusaki1993,Yue1994,nayak99,Lal2002,chen02,Oshikawa2006,Aristov2009,Aristov2010,hou12}. The linear-response properties of $N$-wire junctions within the Tomonaga-Luttinger liquid (TLL) model \cite{Tomonaga1950,Luttinger1963,Giamarchi2004}, which is a powerful formalism for studying interacting electrons in one dimension, are well understood, possibly with the exception \cite{Oshikawa2006,Aristov2013} of the limit of strong attraction for $N\geq 3$. On the contrary, out-of-equilibrium transport through $N$-lead junctions is still a challenging problem. While for two-lead junctions the scaling of the conductance with bias voltage is essentially carried over from the linear-response scaling with temperature or length \cite{Kane1992,Furusaki1996,Sassetti1996,Egger2000,Dolcini2003,Metzner2012}, such a simple connection cannot be made for multi-lead junctions ($N\geq 3$) \cite{Aristov2017,Aristov2018}.

Much work devoted to TLL junctions has been relying on the bosonization method, i.e., a representation of the system in terms of density excitations. This leaves the question of how the nonequilibrium fermionic quasiparticle excitations from the charge reservoirs are transmitted through the junction---and the related question of the determination of the contact resistance---not straightforwardly answered (see Refs.~\cite{Safi1995,maslov95,ponomarenko95,Furusaki1996,alekseev96,egger96,chamon97,gutman10} for $N=2$; for recipes for how to incorporate the contact resistance for $N\geq 3$ see Refs.~\cite{chen02,Oshikawa2006,hou12}). An alternative method, based on a purely fermionic representation \cite{Yue1994,Lal2002,Aristov2009,Aristov2010,Aristov2014,Shi2016}, including the renormalization-group treatment of strong interactions \cite{Aristov2009,Aristov2014,Shi2016}, avoids this problem.

In this paper, we consider charge transport through a multi-lead junction connecting TLL wires in the nonlinear regime. Rather than customarily focusing on the scale dependent (``logarithmic") interaction-induced contributions to the currents, which represent virtual excitation processes (screening), we consider the complementary {\it real} processes. Our purpose here is to demonstrate that the latter give rise to a quite unusual current response normally encountered in the presence of a magnetic field, or certain geometric asymmetry for that matter, in the system.

\begin{figure}[t]
\includegraphics[width=0.95\columnwidth]{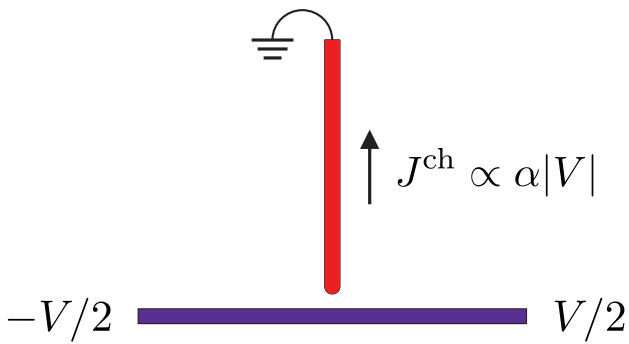}
\caption{Current in the main wire (blue) connected symmetrically (``left vs right") to the tip wire (red) and biased symmetrically with respect to it (grounded tip wire with voltages $\pm V/2$ applied to the main-wire terminals) generates a chiral current $J^{\rm ch}$ in the tip wire, proportional to the modulus of the bias voltage $|V|$ (at zero temperature) and the strength of electron-electron interaction $\alpha$ in the main wire.}
\label{f1}
\end{figure}

Perhaps most surprisingly, we find that in a symmetric Y-junction geometry, i.e., the one with the tip wire attached {\it symmetrically} to the main wire, and for the main wire biased {\it symmetrically} with respect to the grounded tip (Fig.~\ref{f1}), a current through the main wire drives a ``chiral" current in the tip wire. We emphasize that in the absence of interactions---and even when the interaction-induced renormalization of the junction parameters is taken into account---the current in the tip wire is exactly zero under these conditions.

Why the current induced in the tip wire by real interaction processes is chiral: this is because its direction does not depend on the direction of the current in the main wire \cite{remark3}. The direction of the chiral current only depends on the ``sign of interaction" (repulsive vs attractive) in the main wire and the properties of the $S$ matrix that characterizes the junction in the noninteracting limit. Formulated in more general terms, the bias voltage applied to the main wire breaks parity and/or time-reversal symmetries of the differential conductance matrix (in space of the reservoir indices), even if these are preserved at equilibrium. Each of the corresponding interaction-induced terms in the conductance matrix is proportional, at zero temperature, to the sign of one of the voltage differences. That is, this ``emergent chirality" is an essentially nonequilibrium phenomenon, nonexistent in the linear response limit.

As a matter of fact, the combined effect of the bias voltage and interactions on the conductance is equivalent to that of a (local) magnetic flux and/or dipole-like electric field added to the noninteracting junction, which break, respectively, time-reversal and parity symmetries. For the Y junction, this leads to the emergence of the off-diagonal elements of the $2\times 2$ matrix of the ``fundamental" conductances \cite{aristov11,Aristov2017}. Specifically, the antisymmetric and symmetric off-diagonal elements describe an effective Hall and ``photogalvanic" (broadly understood, e.g., in the spirit of Ref.~\cite{belinicher80}) response, respectively \cite{zyuzin04}.

The nonequilibrium symmetry breaking we discuss here also implies a rectification \cite{feldman} of (a part of) the current in the tip wire when the main wire is biased by {\it ac} voltage. In particular, in the symmetric setup of Fig.~\ref{f1}, when the chiral current is the only current in the tip wire, the junction performs as an ideal ``full-wave rectifier" which rectifies a sinusoidal driving of the main wire by generating a {\it dc} current and a double-frequency current, with equal amplitudes, in the tip wire.

Physics behind the emergent chirality is most clearly elucidated by looking at scattering of electrons off nonequilibrium Friedel oscillations of the electron density around the junction. As already mentioned above, the nonequilibrium breaking of time-reversal and parity symmetries that we consider in this paper comes from real interaction-induced processes, as opposed to virtual processes. We formalize our approach to studying the real processes within two complementary frameworks: by directly calculating the currents produced by scattering off Friedel oscillations and by calculating them within the Keldysh formalism.

The crucial ingredient of our approach to the nonequilibrium problem is the recognition of a key difference between the real and virtual processes from the point of view of symmetry which includes chirality. Specifically, let the junction be time-reversal and parity symmetric in the noninteracting limit. In the linear response, taking the renormalization of the junction due to virtual processes into account leaves these symmetries intact. At nonequilibrium, however, the renormalization generically breaks both time-reversal and parity symmetries \cite{Aristov2017}, similarly, in this respect, to the effect of real processes considered here. What makes the concept of emergent chirality precise is that the breaking of time-reversal and parity symmetries in this phenomenon is inherently linked to the chirality of the current. This is in stark contrast to virtual processes for which all currents change their signs---and only signs---when the polarity of {\it all} voltages is changed to the opposite. The current in wire 3 in Fig.~\ref{f1} arises precisely because it breaks time-reversal and/or parity symmetries {\it and} is chiral.

The paper is organized as follows. In Sec.~\ref{s2}, we formulate the model. In Sec.~\ref{s3}, we derive the interaction-induced corrections to the $S$ matrix for the Y junction and the resulting currents to first order in interaction in terms of scattering off nonequilibrium Friedel oscillations. In Sec.~\ref{s4}, we calculate the currents in the $N$-lead junction to first order in interaction within the Keldysh technique. In Sec.~\ref{s5}, we address the symmetry properties of the conductance matrix in the nonlinear response. In Sec.~\ref{s6}, we discuss the TLL renormalization of the parameters of the junction in the context of emergent chirality. In Sec.~\ref{s7}, we analyze the connection of the interaction-induced chiral current to the currents in a noninteracting junction with broken symmetries, in particular, to the Hall current induced by a magnetic flux threading the noninteracting junction. Section \ref{s8} summarizes the results.

\section{Model}
\label{s2}

We consider a junction of $N$ TLL wires labeled by $j=1,2,\ldots,N$, each connected to a reservoir of electrons characterized by a fermionic distribution function $f_j(\epsilon )$, where $\epsilon$ is the energy of electrons emitted from reservoir $j$. For the most part of the paper, we focus on the case of
thermal reservoirs characterized by distinct chemical potentials $\mu_j$ and the same temperature $T$, i.e., by $f_j(\epsilon)=1/\{\exp [(\epsilon -\mu_j)/T]+1\}$. This corresponds to the B{\"u}ttiker-Landauer formalism with ideal terminals, defined as absorbing anything incident on them and emitting electrons with the equilibrium distribution functions that are independent of the state of the system connected to the terminals.

We use a fermionic representation for the scattering problem.
% \cite{Yue1994,Lal2002,Aristov2010,Aristov2014}.
% which allows us to avoid the problems related to the contact resistance in the bosonic approach.
The electronic states in each wire are separated into incoming and outgoing components, labeled by the chirality index $\eta_{j}=\pm $, with the ``right-moving" waves ($\eta _{j}=+$) running away from the junction. The coordinate $x$ in each of the wires varies from $x=0$ to $x=L$, where $L$ is the length of the wire from the junction to the reservoir. Assuming a linearized dispersion of spinless (spin-polarized) electrons, the Hamiltonian in wire $j$ reads ($\hbar=1$)
\be
H_j=\sum_{\eta_j=\pm}\int_0^L\!dx\,v_j\left(-i\eta_j\Psi_{\eta_j}^\dagger\partial_{x}\Psi_{\eta_j}+\pi\alpha_jn_{\eta_j}n_{-{\eta_j}}\right)~.
\label{1}
\ee
Here $n_{\eta_j}=\Psi_{\eta_j}^\dagger\Psi_{\eta_j}$ is the chiral density, $v_j>0$ is the electron velocity (corrected by the interaction between electrons of the same chirality), and $\alpha_j$ characterizes the strength of the (short-ranged) interaction, given by the difference of forward- and backward-scattering Fourier components of the interaction potential. For simplicity, we assume equal Fermi velocities $v_j=v$ in all wires (but allow for the difference between $\alpha_j$ for different $j$). The wires are connected to each other at the origin by a symmetric single-particle $S$ matrix $S_{jj^\prime}$.

It is worth emphasizing that the above model for the wires and their junction, assuming that the $S$ matrix is symmetric, respects time-reversal symmetry on the Hamiltonian level, which translates into the property of the conductance matrix being symmetric in the linear response limit. Time-reversal symmetry in the conductance matrix is only broken, then, in the nonlinear response, as was already mentioned in Sec.~\ref{s1}. If we assume that the Hamiltonians (\ref{1}) for different $j$ and the $S$ matrix at the junction possess, additionally, 1$\leftrightarrow$2 parity symmetry, this symmetry in the conductance matrix will also only be violated beyond the linear response, similarly to time-reversal symmetry.

\section{Triple junction out of equilibrium: Emergent chirality}
\label{s3}

We begin by considering a symmetric Y junction (also referred to as a triple junction) consisting of the main wire (leads $1,2$) and the tunneling tip (lead $3$). The $S$ matrix for this setup is given by
\be
\hat{S}=
\begin{pmatrix}
r & t & t_3 \\
t & r & t_3 \\
t_3 & t_3 & r_3
\end{pmatrix}
\label{S33-matrix}
\ee
in the basis of terminals $1,2,3$. Here $r$ and $t$ denote the reflection and transmission amplitudes of the junction within the main wire, respectively, and $r_3$ and $t_3$ are the reflection and transmission amplitudes for the tip. Time-reversal symmetry means that the $S$ matrix is symmetric (in line with the comment at the end of Sec.~\ref{s2}) and 1$\leftrightarrow$2 parity symmetry means that, in addition, $S_{13}=S_{23}$. The scattering amplitudes in Eq.~(\ref{S33-matrix}) can be expressed in terms of three angles, $\theta$, $\psi$, and $\gamma$ (up to an unobservable global phase) as
\bea
r&=&\frac{1}{2}\left(\cos\theta+e^{-i\psi}\right)e^{i\gamma}~,\quad r_3=\cos\theta e^{-i\gamma}~,  \notag \\
t&=&\frac{1}{2}\left(\cos\theta-e^{-i\psi}\right)e^{i\gamma}~,\quad t_3=\frac{i}{\sqrt{2}}\sin\theta~.
\label{rr3tt3}
\eea

The phase $\gamma$ drops out from the conductance matrix of the noninteracting junction and does not affect the interaction-induced renormalization of the other two phases \cite{aristov11}. As will be shown below, $\gamma$ also drops out from the nonequilibrium chiral current. It is worth noting that the angle $\psi$ is zero in the model of a local tunneling tip (lead $3$ connected to the main wire at a single point). As we will see, the nonequilibrium-induced chirality is inherently related to $\psi\neq 0$.

In Secs.~\ref{s3a} and \ref{s3b}, we start with the case in which the interaction strength is the same in half-wires $1$ and $2$, i.e., along the main wire, $\alpha_1=\alpha_2=\alpha$, whereas the tip is noninteracting, $\alpha_3=0$. Our prime goal here is to calculate the current in wire 3, $J_3$, in a way that is more transparent, both physically and mathematically, than the Keldysh formalism presented later in Sec.~\ref{s4}. As mentioned in Sec.~\ref{s1}, this is achieved by studying scattering off nonequilibrium Friedel oscillations. We show in Sec.~\ref{s3b}, to leading order in $\alpha$, that the current $J_3$ is even in the voltage $V$ applied symmetrically to the main wire, with $J_3\propto\alpha |V|$ for zero temperature. In Sec.~\ref{s3c}, we generalize the approach of Secs.~\ref{s3a} and \ref{s3b} to allow for different $\alpha_j$ in different wires and for an arbitrary distribution of voltages.

\subsection{Scattering off nonequilibrium Friedel oscillations}
\label{s3a}

In the spirit of Ref.~\cite{Yue1994}, we first account for interaction in the main wire perturbatively through the inclusion of additional scattering off Friedel oscillations around the junction. This process involves the Hartree interaction potential characterized by the backscattering (Hartree) interaction constant $\alpha_{H}$. The total (Hartree plus exchange) interaction-induced correction to the $S$ matrix (\ref{S33-matrix}) is obtainable by the replacement $\alpha_{H}\rightarrow -\alpha$ in the Hartree correction.

\begin{figure}[t]
\includegraphics[width=0.95\columnwidth]{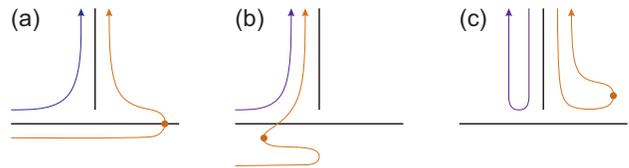}
\caption{Three types of processes contributing to the current $J_3$ in the tip. Two interfering waves in each of the processes are shown by blue (``bare'') and red (``interacting''). The red dots denote backscattering off the Friedel oscillations within the main wire. Each process is complemented by its mirror (left-right) image.}
\label{f2}
\end{figure}

Friedel oscillations in the main wire give rise to the quantum interference of scattered waves in three types of scattering processes denoted as (a), (b), and (c) in Fig.~\ref{f2}. Importantly, the Friedel oscillations in wires 1 and 2 are created by electrons supplied by their ``own" terminals 1 and 2, respectively, being governed by the distribution functions $f_{1,2}$ in the corresponding reservoirs. In particular, for $\mu_1\neq\mu_2$, the Friedel oscillations to the left and to the right of the junction have different periods, which will lead to a peculiar behavior of the currents.

It is convenient to change $x\to -x$ in wire 1, so that $x$ varies from $-\infty$ to $+\infty$ in the infinite main wire $1+2$ [in Sec.~\ref{s3}, we take the limit $L\to\infty$, with $L$ from Eq.~(\ref{1}), from the very beginning]. It is also convenient to count the energies of electrons from the common bottom for all wires. The oscillatory part of the Hartree potential in the main wire reads:
\be
U_H(x)=\alpha_H\,2\text{Re}\!\int_0^\infty\!\frac{d\epsilon^{\prime}}{2\pi}e^{-2i\epsilon^{\prime}x/v}\times
\begin{cases}
r\ f_1(\epsilon^{\prime }), & \ x<0~, \\
r^* f_2(\epsilon^{\prime }), & \ x>0~.
\end{cases}
\label{VH}
\ee

In process (a), the wave emitted from terminal 1 at energy $\epsilon$ is first transmitted through the junction, with the amplitude $t$, into wire 2. Next, it is backscattered off the Friedel oscillation (red dot in Fig.~\ref{f2}a) with the reflection amplitude determined by the matrix element of the potential $U_H(x)$ [Eq.~(\ref{VH})]. The Friedel oscillation for $x>0$ is produced by electrons that are emitted from terminal 2 with the distribution function $f_2(\epsilon^{\prime})$ and then reflected from the junction with the amplitude $r^*$. Finally, the wave leaves the main wire and escapes into wire 3 with the amplitude $t_3$. The corresponding correction to the transmission amplitude $t_3$ at energy $\epsilon$ is given by
\be
\delta t_3^a(\epsilon)=\frac{\alpha_H}{2} r^*tt_3\int_0^\infty\!d\epsilon^{\prime }\,f_2(\epsilon^{\prime })\,\frac{1}{\epsilon-\epsilon^{\prime}+i0}~.
\label{dt3a}
\ee
The energy denominator in Eq.~(\ref{dt3a}) appears as the result of the integration over the position of the point at which the scattering off the Friedel oscillation occurs.

The contribution of process (b) to the transmission amplitude $t_3$ is obtainable from Eq.~(\ref{dt3a}) by changing $t\to r$ and $f_2\to f_1$:
\be
\delta t_3^b(\epsilon)=\frac{\alpha_H}{2}r^*rt_3\int_{0}^\infty\!d\epsilon^{\prime}\,f_1(\epsilon^{\prime})\,\frac{1}{\epsilon-\epsilon^{\prime}+i0}~.
\label{dt3b}
\ee
Indeed, interaction-induced scattering at $x<0$ requires reflection from the junction with the amplitude $r$ while the Friedel oscillation in wire 1 is produced by electrons emitted from terminal 1 with the distribution function $f_1(\epsilon^{\prime })$.

Similarly, the contribution of process (c), which is scattering of the wave emitted from terminal 3 off the Friedel oscillation in wire 2, is obtainable from Eq.~(\ref{dt3a}) by changing $t\to t_3$:
\be
\delta r_3^c(\epsilon)=\frac{\alpha_H}{2}r^*t_3t_3\int_0^\infty\!d\epsilon^{\prime }\,f_2(\epsilon^{\prime })\,\frac{1}{\epsilon-\epsilon^{\prime}+i0}~.
\label{dr3c}
\ee
Finally, the contributions of the right-left ``mirror images'' of processes (a), (b) to $\delta t_3(\epsilon)$ and process (c) to $\delta r_3(\epsilon)$ are obtainable from Eqs.~(\ref{dt3a}), (\ref{dt3b}), and (\ref{dr3c}), respectively, by exchanging $f_1\leftrightarrow f_2$.

The principal-value integrals over energy in Eqs.~(\ref{dt3a})-(\ref{dr3c}) produce a logarithmically singular correction to the $S$ matrix, which can be further accounted for by a renormalization-group summation \cite{Aristov2010,Aristov2017}. This gives rise to the currents that are odd in the voltages. Remarkably, it is the pole contribution to the integrals (\ref{dt3a})-(\ref{dr3c}) that, while not producing singular corrections to the scattering amplitudes, leads to an even-in-voltage current---the emergence of which is the main prediction of this work.

\subsection{Chiral current}
\label{s3b}

To leading order in the interaction-induced scattering, the correction to the charge current of noninteracting electrons ($e>0$)
\begin{align}
J_{3}^{(0)}=-e\!\int_0^\infty\!\!\frac{d\epsilon}{2\pi}&\left\{\left[\,f_{1}(\epsilon)+f_{2}(\epsilon)\,\right]|t_{3}(\epsilon)|^2\right.\notag\\
+ &\left. f_{3}(\epsilon)\left(|r_{3}|^{2}-1\right)\!\right\}
\label{J3}
\end{align}
in wire 3 is given by
\begin{align}
\delta J_{3}\!=\!-e\,2\text{Re}\int_0^\infty\!\!\frac{d\epsilon}{2\pi}&\left\{f_1(\epsilon)t_3^\ast\left[ \delta t_3^a(\epsilon)+
\delta t_3^b(\epsilon )\right] \right.\notag\\
+ &\left. f_3(\epsilon)r_3^\ast\delta r_3(\epsilon)\right\}+(f_1\leftrightarrow f_2)~,
\label{dJ3}
\end{align}
where the exchange $f_{1}\leftrightarrow f_{2}$ should be performed everywhere, both in the factor in Eq.~(\ref{dJ3}) and in Eqs.~(\ref{dt3a}), (\ref{dt3b}), and (\ref{dr3c}) for the corrections to the scattering amplitudes. Note that only the distribution functions supplied by the terminals enter the expression for the current. This not only concerns the explicit factor in Eq.~(\ref{dJ3}) but also the corrections $t_{3}$ and $r_{3}$ determined by the Friedel oscillations. In Eqs.~(\ref{J3}) and (\ref{dJ3})---and everywhere below, for each of wires 1,2,3---the charge currents are defined as positive when running in the direction away from the junction.

We now focus on the simplest and perhaps most interesting case when the main wire is biased symmetrically (Fig.~\ref{f1}): $\mu_1-\mu_3=-V/2$, $\mu_2-\mu_3=V/2$, and $\mu_3=\Lambda$ with an arbitrary $\Lambda$ counted from the band bottom. According to Eq.~(\ref{J3}), in the absence of interaction, $\alpha=0$, the current in wire 3 is zero for this distribution of voltages. It follows from the structure of Eq.~(\ref{dJ3}) that, under the same conditions, $\delta J_3$ (and hence the total current in the presence of interaction) is an even function of $V$.

It is convenient to express the current in terms of the integrals
\be
I_{kl}=\int_0^\infty\!\frac{d \epsilon}{2\pi}\,f_k(\epsilon)\int_0^\infty\!d \epsilon^{\prime }f_l(\epsilon^{\prime })\,\frac{1}{\epsilon-\epsilon^{\prime }+i0}~.
\label{10}
\ee
Including the Fock contribution, as discussed above, by replacing $\alpha_H\to -\alpha$, we write:
\begin{align}
J_3=e\alpha\,\text{Re}&\left\{|t_3|^2r^*t(I_{12}+I_{21})+|t_3|^2|r|^2(I_{11}+I_{22})\right. \notag \\
&+\left. t_3^2r^*_3r^*(I_{32}+I_{31})\right\}~.
\label{J3Ikl}
\end{align}
By exchanging $\epsilon\leftrightarrow \epsilon^{\prime }$, the principal values of the integrals $I_{11}$ and $I_{22}$ vanish, $\text{p.v.}\{I_{11}\}=\text{p.v.}\{I_{22}\}=0$. Thus, process (b) and its mirror process ($I_{11}$ and $I_{22}$, respectively) do not contribute to $J_3$. The principal values of the integrals $I_{12}$ and $I_{21}$ vanish in the combination $\text{p.v.}\{I_{12}+I_{21}\}=0$. The principal value of the combination $I_{32}+I_{31}$ does not vanish, but is small in $1/\Lambda$. Thus, in the limit $\{V,T\}/\Lambda\to 0$, the current $J_3$ is entirely determined by the pole contributions to $I_{kl}$.

For clarity, in the remainder of Sec.~\ref{s3b}, we focus on the zero-temperature case, with
\bea
\text{Im} I_{12}\!=\!\text{Im} I_{21}\!\!&=&\!\!-\frac{1}{2} \int_0^\infty\!d\epsilon f_1(\epsilon) f_2(\epsilon) \!=\!-\frac{1}{2}
\left(\Lambda-\frac{|V|}{2}\right)~,  \notag \\
\text{Im}\{I_{32}+I_{31}\}\!\!&=&\!\!-\frac{1}{2} \int_0^\infty\!d\epsilon f_3(\epsilon)\left[f_1(\epsilon)+f_2(\epsilon)\right]  \notag \\
&=&\!-\frac{1}{2} \left(2\Lambda-\frac{|V|}{2}\right)~.
\label{ImI}
\eea
Substituting these results into Eq.~(\ref{J3Ikl}), we observe that the terms proportional to $\Lambda$ (equilibrium current) cancel out because of the unitarity of the $S$ matrix [specifically, the cancellation can be seen by multiplying the unitarity condition $t_3^*t+t_3^*r+r_3^*t_3=0$ by $t_3r^*$ and taking the imaginary part of the product, which gives ${\text{Im}\{|t_3|^2t r^*+t_3^2 r_3^* r^*\}=0}$]. The nonequilibrium current is, however, finite and proportional to $|V|$:
\be
J_3=-\frac{e\alpha |V|}{4}\,\text{Im}\left\{2|t_3|^2t r^*+t_3^2 r_3^* r^*\right\}~,
\label{13}
\ee
where, by unitarity, the contribution of process (a) and its mirror process (combined they give the first term in the brackets) is $(-2)$ times the contribution of process (c) and its mirror process (the second term), i.e.,
\be
J_3=-\frac{e \alpha |V|}{4}\,\text{Im}\left\{|t_3|^2t r^*\right\}~.
\label{J3V}
\ee

Using the parametrization (\ref{rr3tt3}), $J_3$ is expressed in terms of the angles $\theta$ and $\psi$ as
\be
J_3=-\frac{e\alpha |V|}{16}\sin^2\theta\cos\theta\sin\psi~.
\label{J3thetapsi}
\ee
Note that, as already mentioned at the beginning of Sec.~\ref{s3}, the phase $\gamma$ [Eq.~(\ref{rr3tt3})] does not enter the induced current in wire 3. Equation (\ref{J3thetapsi}) shows that $J_3$ is zero for the decoupled ($t_3=0$, i.e., $\theta=0$ or $\pi$ for arbitrary $\psi$) or perfectly absorbing ($r_3=0$, i.e., $\theta=\pi/2$ or $3\pi/2$ for arbitrary $\psi$) main wire, or pointlike coupling between the main wire and the tip ($\psi=0$).

The ``picture'' of elastic scattering off Friedel oscillations is particularly instructive in that it clearly demonstrates the meaning of the energy integration in Eqs.~(\ref{dt3a}), (\ref{dt3b}), and (\ref{dr3c}) for the corrections to the scattering amplitudes with the biased distribution functions. Taking the pole terms in the interaction-induced scattering amplitudes, which produced the current in Eq.~(\ref{J3thetapsi}), is a hallmark of real processes, as opposed to screening. The latter involves integration over the energies of virtual excitations and results in the renormalization of the scattering amplitudes. The peculiarity of $J_3$ for the symmetric distribution of voltages specified above Eq.~(\ref{10}) is that the principal-value terms in the scattering amplitudes cancel out, so that the interaction-induced current is solely determined by the pole terms.

\subsection{Interaction-modified $S$ matrix and the currents}
\label{s3c}

The  chiral current is thus seen to come from elastic scattering off Friedel oscillations, which can be formalized in terms of the ``pole-related" correction $\delta S_{jk}^{(p)}(\epsilon)$ to the single-particle $S$ matrix at energy $\epsilon$. Extending the calculation in Sec.~\ref{s3b} to arbitrary $\alpha_{1,2,3}$, we also see that $\alpha_j$ appears only in the combination $\alpha_jf_j(\epsilon)$. Specifically,
\bea
\hspace{-5mm}\delta\hat{S}^{(p)}(\epsilon)\!\!&=&\!\!\frac{i}{2}\pi\alpha_1f_1(\epsilon)\!\left(
\begin{array}{lll}
|r|^2r & |r|^2t & |r|^2t_3 \\
|r|^2t & t^2r^* & tr^*t_3 \\
|r|^2t_3 & tr^*t_3 & t_3^2r^*
\end{array}
\right) \notag\\
&+&\!\!\frac{i}{2}\pi\alpha_2f_2(\epsilon)\!\left(
\begin{array}{lll}
t^2r^* & |r|^2t & tr^*t_3 \\
|r|^2t & |r|^2r & |r|^2t_3 \\
tr^*t_3 & |r|^2t_3 & t_3^2r^*
\end{array}
\right)
\notag\\
&+&\!\!\frac{i}{2}\pi\alpha_3f_3(\epsilon)\!\left(
\begin{array}{lll}
t_3^2r_3^* & t_3^2r_3^* & |r_3|^2t_3 \\
t_3^2r_3^* & t_3^2r_3^* & |r_3|^2t_3 \\
|r_3|^2t_3 & |r_3|^2t_3 & |r_3|^2r_3
\end{array}
\right).
\label{16}
\eea
We emphasize that the $S$ matrix for given $\epsilon$ retains its time-reversal symmetric form for arbitrary $f_{1,2,3}(\epsilon)$. At the same time, 1$\leftrightarrow$2 parity symmetry of the $S$ matrix is only preserved if $\alpha_1f_1(\epsilon)=\alpha_2f_2(\epsilon)$.

The pole-related $\mathcal{O}(\alpha_j)$ correction to the noninteracting currents $J_j^{(0)}$ is given in terms of $\delta\hat{S}^{(p)}(\epsilon)$ by
\be
\delta J_j=-e\int_0^\infty\!\frac{d\epsilon}{2\pi}\sum_kA_{jk}(\epsilon)f_k(\epsilon)~,
\label{Jip}
\ee
where
\be
A_{jk}(\epsilon)=2\text{Re}\left\{S_{jk}^\ast\delta S_{jk}^{(p)}(\epsilon)\right\}~.
\label{18}
\ee
From Eqs.~(\ref{S33-matrix}) and (\ref{16}), we have for the matrix $\hat{A}(\epsilon)$:
\bea
&&\hat{A}(\epsilon) = \frac{\pi}{4}\sin^{2}\theta\cos\theta\sin\psi \notag \\
&&\times\left[\,\alpha_1f_1(\epsilon)
\begin{pmatrix}
0 & 0 & 0 \\
0 & 1 & -1 \\
0 & -1 & 1
\end{pmatrix}
+\alpha_2f_2(\epsilon)\begin{pmatrix}
1 & 0 & -1 \\
0 & 0 & 0 \\
-1 & 0 & 1
\end{pmatrix}
\right.\notag\\
&&\left.\hspace{2.5mm}+\,\alpha_3f_3(\epsilon)\begin{pmatrix}
1 & -1 & 0 \\
-1 & 1 & 0 \\
0 & 0 & 0
\end{pmatrix}
\,\right]~.
\label{Mij}
\eea
Despite the somewhat cumbersome form of Eq.~(\ref{16}), all the entries to the matrix $\hat{A}(\epsilon)$ are proportional to the single parameter
\be
\mathcal{A}=\frac{1}{4}\sin^{2}\theta\cos\theta\sin\psi~,
\label{caligra}
\ee
already encountered in Eq.~(\ref{J3thetapsi}). The matrix $\hat{A}(\epsilon)$ is still time-reversal symmetric, similarly to $\delta\hat{S}^{(p)}(\epsilon)$ [Eq.~(\ref{16})]. It is also worth remarking that the product $\alpha_lf_l$ appears in the matrix elements $A_{jk}(\epsilon)$ with $j,k\neq l$. This prevents the emergence of terms with $f_l^2(\epsilon)$ in the current, as it should be.

Substituting Eq.~(\ref{Mij}) in Eq.~(\ref{Jip}), we obtain
\begin{align}
&\begin{pmatrix}
\delta J_1 \\ \delta J_2 \\ \delta J_3
\end{pmatrix}
=
\frac{e}{2}{\cal A}\notag\\
&\times\begin{bmatrix}
\alpha_2\left(\left\langle f_2f_3\right\rangle-\left\langle f_1f_2\right\rangle\right)+
\alpha_3\left(\left\langle f_2f_3\right\rangle-\left\langle f_1f_3\right\rangle\right)\\
\alpha_3\left(\left\langle f_1f_3\right\rangle-\left\langle f_2f_3\right\rangle\right)+
\alpha_1\left(\left\langle f_1f_3\right\rangle-\left\langle f_1f_2\right\rangle\right)\\
\alpha_1\left(\left\langle f_1f_2\right\rangle-\left\langle f_1f_3\right\rangle\right)+
\alpha_2\left(\left\langle f_1f_2\right\rangle-\left\langle f_2f_3\right\rangle\right)
\end{bmatrix}~,
\label{21}
\end{align}
where
\begin{align}
\langle f_jf_k\rangle=\int_0^\infty\!\!d\epsilon\,f_j(\epsilon)f_k(\epsilon)~.
\label{22}
\end{align}
Note that $\delta J_j$ does not depend on $\alpha_j$; in particular, the current in the tip wire only depends on the interactions in the main wire.

For $T=0$, the average in Eq.~(\ref{22}) is written as
\be
\langle f_jf_k\rangle=\min\{\mu_j,\mu_k\}=\frac{1}{2}(\mu_j+\mu_k-|\mu_j-\mu_k|)~.
\label{23}
\ee
The (differential) conductance matrix $\delta\hat{G}$, which relates $\delta J_{1,2,3}$ and $\mu_{1,2,3}$ by means of $\delta G_{jk}=e\,\partial\delta J_j/\partial\mu_k$, is then representable as a sum of two terms:
\be
\delta\hat{G}=\hat{G}^{\rm reg}+\hat{G}^{\rm ch}~,
\label{24}
\ee
where the ``regular" part $\hat{G}^{\rm reg}$ does not depend on the bias voltages and the chiral part $\hat{G}^{\rm ch}$ depends on their signs (and, for $T=0$, on their signs only). We have:
\be
\hat{G}^{\rm reg}=-\frac{e^2}{4}{\cal A}
\begin{pmatrix}
\alpha_2+\alpha_3  & -\alpha_3 & -\alpha_2 \\
-\alpha_3 & \alpha_1+\alpha_3 & -\alpha_1 \\
-\alpha_2 & -\alpha_1 & \alpha_1+\alpha_2
\end{pmatrix}
\label{25}
\ee
and
\begin{widetext}
\be
\hat{G}^{\rm ch}=\frac{e^2}{4}{\cal A}
\begin{bmatrix}
\alpha_2s_{12}-\alpha_3s_{31} & -\alpha_2(s_{12}+s_{23})-\alpha_3s_{23} & \alpha_2s_{23}+\alpha_3(s_{23}+s_{31}) \\
\alpha_1(s_{12}+s_{31})+\alpha_3s_{31} & -\alpha_1s_{12}+\alpha_3s_{23} & -\alpha_1s_{31}-\alpha_3(s_{23}+s_{31}) \\
-\alpha_1(s_{12}+s_{31})-\alpha_2s_{12} & \alpha_1s_{12}+\alpha_2(s_{12}+s_{23}) & \alpha_1s_{31}-\alpha_2s_{23}
\end{bmatrix}~,
\label{26}
\ee
\vspace{6mm}
\end{widetext}
where $s_{jk}={\rm sgn}(\mu_j-\mu_k)$. Importantly, while $\hat{G}^{\rm reg}$ remains symmetric, i.e., respects time-reversal symmetry, this is generically not the case for $\hat{G}^{\rm ch}$. Nor does $\hat{G}^{\rm ch}$ maintain 1$\leftrightarrow$2 parity symmetry even for $\alpha_1=\alpha_2$, again, in contrast to $\hat{G}^{\rm reg}$. Breaking of time-reversal symmetry in the chiral part of the conductance matrix is in contrast to its maintainance in the $S$ matrix at given $\epsilon$ [Eq.~(\ref{16})].

One of the remarkable properties of the sum of the matrices $\hat{G}^{\rm reg}$ and $\hat{G}^{\rm ch}$ is that the current $\delta J_i$ does not depend on the largest of the chemical potentials $\mu_{1,2,3}$, i.e., only depends on the difference of the two smallest ones. A corollary is that $\delta J_i=0$ when the two smallest chemical potentials are degenerate.

Another point to notice is that the conductance matrix $\hat{G}$ in general, and $\hat{G}^{\rm ch}$ in particular, is characterized by redundancy, because it obeys two constraints: $\sum_jG_{jk}=0$ (charge conservation, or Kirchhoff's current law for that matter) and $\sum_kG_{jk}=0$ (invariance under an arbitrary shift of the reference point for the chemical potentials, or Kirchhoff's voltage law). With these constraints, the most generic structure of the conductance matrix for a Y junction is parametrized by three constants $\xi_{1,2,3}$:
\be
\hat{G}\propto
\begin{pmatrix}
1 & -\xi_1 & \xi_1-1 \\
-\xi_2 & \xi_3 & \xi_2-\xi_3 \\
\xi_2-1 & \,\,\xi_1-\xi_3\,\, & 1-\xi_1-\xi_2+\xi_3
\end{pmatrix}~,
\label{27}
\ee
up to a common multiplier. The difference between $\xi_1$ and $\xi_2$ signifies broken time-reversal symmetry. For $\xi_1=\xi_2$, the difference between $\xi_3$ and 1 breaks 1$\leftrightarrow$2 parity and the difference between $\xi_1$ and 1/2 breaks 2$\leftrightarrow$3 parity. Equation (\ref{27}) thus shows that no symmetry that can possibly be broken is left intact, for a generic distribution of the chemical potentials $\mu_{1,2,3}$ and a generic set of the interaction constants $\alpha_{1,2,3}$, in $\hat{G}^{\rm ch}$---despite the noninteracting $S$ matrix from Eq.~(\ref{S33-matrix}) being highly symmetric. We will further discuss the symmetry properties of the conductance matrix from the point of view of emergent chirality in Sec.~\ref{s5}.

\section{Emergent chirality from the Keldysh formalism}
\label{s4}

Having clarified the origin of the chiral current in a Y junction in terms of the interaction-induced corrections to the single-particle $S$ matrix, we proceed with the analysis of the general case of $N$-lead TLL junctions within the fermionic Keldysh formalism.

\subsection{Fermionic Keldysh technique for an $N$-lead junction}
\label{s4a}

In the absence of interaction, the net current in wire $j$ can be written as the difference of the ``in" and ``out" currents,
\be
J_j^{(0)}=-\frac{e}{2}\int \frac{d\epsilon}{2\pi }\sum_k\left(\delta_{jk}-|S_{jk}|^{2}\right)h_k(\epsilon)~,
\label{netJ0}
\ee
with the partial contribution of wire $l$ weighted with the ``Keldysh function" $h_k(\epsilon)=1-2f_k(\epsilon)$ [cf.\ Eq.~(\ref{J3})]. In the case of thermal reservoirs at temperature $T$, on which we concentrate in this paper,
\be
h_k(\epsilon)=\tanh\frac{\epsilon-\mu_k}{2T}~.
\label{29}
\ee

We now derive the contribution to the currents to first order in interaction within the Keldysh technique, largely following the formulation of a similar nonequilibrium problem for the case of a two-lead junction in Ref.~\cite{Aristov2014}. The interaction-induced current in wire $j$ at position $z$ to first order in $\alpha_l$ is given, within the TLL model, by
\be
\delta J_j(z)=-e\!\int\!\frac{d\epsilon}{2\pi}\!\int\!\frac{d\omega}{2\pi}\!\int_0^L\!\!dx\sum_{l}(2\pi i\alpha_{l}v)\,T_{jl}(z,x;\epsilon,\omega)~,
\label{current}
\ee
where $T_{jl}$ represents a ``triangle'' of the noninteracting Green's functions:
\begin{align}
&\!\!T_{jl}(z,x;\epsilon,\omega)=v\!\!\sum_{\mu=1,2}\sum_{\eta_l=\pm }\mbox{Tr}_K\left[\,\check{\gamma}^\text{ext}
\check{\mathcal{G}}_{\epsilon}(j,+,z\,|\,l,-\eta_{l},x) \right. \notag \\
&\times\left. \check{\bar{\gamma}}^\mu\check{\mathcal{G}}_{\epsilon+\omega }(l,-\eta_l,x\,|\,l,\eta_l,x)\check{\gamma^\mu}
\check{\mathcal{G}}_\epsilon(l,\eta_l,x\,|\,j,+,z)\,\right].
\label{triangle}
\end{align}
The Green's function $\check{\mathcal{G}}_{\epsilon}$ is a $2\times 2$ matrix in Keldysh space (in the Larkin-Ovchinnikov basis),
\be
\check{\mathcal{G}}_\epsilon =
\begin{pmatrix}
{\mathcal{G}}^R_\epsilon  & {\mathcal{G}}^K_\epsilon \\
0 & {\mathcal{G}}^A_\epsilon
\end{pmatrix}~,
\label{32}
\ee
with the arguments of $\check{\cal G}_\epsilon(l,\eta',y\,|\,j,\eta,x)$ denoting propagation with energy $\epsilon$ from point $x$ in wire $j$ to point $y$ in wire $l$, with the initial and final chiralities $\eta$ and $\eta'$, respectively. Scattering off the junction is encoded in $\check{\mathcal{G}}_{\epsilon}$ (with the spatial coordinates $x\neq 0$ and $y\neq 0$) through the $S$-matrix elements $S_{jk}$:
\begin{align}
\check{\mathcal{G}}_\epsilon(l,+,y\,&|\,j,+,x)=-\frac{i}{v}e^{i\epsilon\tau_{++}} \notag\\
&\times\left[
\begin{array}{cc}
\Theta (\tau_{++})\delta_{lj} & \ \sum_mS_{lm}h_m(\epsilon)S_{jm}^\ast \\
0 & \ -\Theta (-\tau_{++})\delta_{lj}
\end{array}
\right], \label{33} \\
\check{\mathcal{G}}_\epsilon(l,+,y\,&|\,j,-,x)=-\frac{i}{v}e^{i\epsilon \tau_{+-}}\left[
\begin{array}{cc}
S_{lj} & S_{lj}h_j(\epsilon) \\
0 & 0
\end{array}
\right], \label{34} \\
\check{\mathcal{G}}_\epsilon(l,-,y\,&|\,j,+,x)=-\frac{i}{v}e^{i\epsilon\tau_{-+}}\left[
\begin{array}{cc}
0 & h_{l}(\epsilon)S_{jl}^{\ast } \\
0 & - S_{jl}^{\ast}
\end{array}
\right], \label{35} \\
\check{\mathcal{G}}_\epsilon(l,-,y\,&|\,j,-,x)=-\frac{i}{v}e^{i\epsilon\tau_{--}} \notag\\
&\times\left[
\begin{array}{cc}
\Theta(\tau_{--})\delta_{lj} & h_l(\epsilon)\delta_{lj} \\
0 & -\Theta(-\tau_{--})\delta_{lj}
\end{array}
\right], \label{36}
\end{align}
where $\tau _{\eta\eta'}=(\eta y-\eta' x)/v$ and $\Theta(\tau)$ is the step function. Unitarity $\hat{S}^{-1}=\hat{S}^\dagger$ is explicitly used in Eqs.~(\ref{33}) and (\ref{35}) for the Green's functions with the initial chirality $\eta_j=+$. The integration over $\epsilon$ is performed with infinite limits even after being put on the mass shell, i.e., the energy $\Lambda$, considered in Sec.~\ref{s3} as finite, is sent to $\infty$ in Eq.~(\ref{current}) from the very beginning.

The trace $\mbox{Tr}_{K}$ in Eq.~(\ref{triangle}) is over the Keldysh indices. The fermion-boson vertices, $\check{\gamma }^{\mu }$ and $\check{\bar{\gamma}}^{\mu}$, and the external (current) vertex $\check{\gamma}^\text{ext}$ are given by
\be
\check\gamma^{1}=\check{\bar{\gamma}}^{2}=\frac{1}{\sqrt{2}}\begin{pmatrix}
1 & 0\\
0 & 1
\end{pmatrix}~,\quad
\check\gamma^{2}=\check{\bar{\gamma}}^{1}=\frac{1}{\sqrt{2}}\begin{pmatrix}
0 & 1 \\
1 & 0
\end{pmatrix}~,
\label{37}
\ee
and
\be
\check{\gamma}^\text{ext}=\frac{i}{2}
\begin{pmatrix}
1 & 1 \\
-1 & -1
\end{pmatrix}~.
\label{38}
\ee
Taking the Keldysh trace, we find that only the ingoing (outgoing) chirality $\eta _{l}=-1$ ($\eta _{l}=1$) contributes to $T_{jl}$ [Eq.~(\ref{triangle})] for $\mu=1$ ($\mu=2$). The result for $T_{jl}$ can be written as
\bea
T_{jl}(z,x;\epsilon,\omega)&=&\frac{i}{2v^2}\sum_m\,{\rm Im}\left\{e^{-2i\omega x/v}B_{jlm}\right\} \notag \\
&\times& [\,h_l(\epsilon+\omega)-h_l(\epsilon )\,]\,h_m(\epsilon)~,
\label{Tjl}
\eea
where we introduce
\be
B_{jlm}=S_{jl}S_{ll}^{\ast}S_{lm}S_{jm}^\ast~.
\label{Ajlm}
\ee
We observe that the position $z$ of the measurement drops out in Eq.~(\ref{Tjl}), as expected for the dc response.

Note that the integration over $\omega$ in Eq.~(\ref{current}) of the $\omega$ independent part of the product $[\,h_l(\epsilon+\omega)-h_l(\epsilon )\,]\,h_m(\epsilon)$
from Eq.~(\ref{Tjl}) produces zero except for the $\delta(x)$ singularity at $x=0$, where Eq.~(\ref{Tjl}) is, as mentioned above Eq.~(\ref{33}), not valid. In fact, the result is zero also for $x=0$. This is because of the general condition, required by causality [the vanishing of the 21 (lower left) matrix element of the fermion self-energy in the basis of Eq.~(\ref{32}) for that matter], that the sum of the retarded and advanced Green's functions with the same arguments, both taken at exactly zero propagation time is zero: $\int\!d\omega\left({\cal G}^R_{\epsilon+\omega}+{\cal G}^A_{\epsilon+\omega}\right)=0$ also for $x=0$. Therefore, the product $h_l(\epsilon)h_m(\epsilon)$ in $T_{ji}$ does not enter any observable. The same is true with regard to the $\omega$ independent term in the product $h_l(\epsilon+\omega)h_m(\epsilon)$. It is convenient, however, to keep them both while integrating $T_{jl}$ over $\epsilon$ and shift the lower limit of the $x$ integration in Eq.~(\ref{current}) to $x=0+$.

We represent the integral over $\epsilon$ of the Keldysh functions from Eq.~(\ref{Tjl}) in the following form [this is where keeping the product $h_l(\epsilon)h_m(\epsilon)$ is useful]:
\begin{align}
\frac{1}{2}\int\!d\epsilon\,[\,h_l(\epsilon+\omega)&-h_l(\epsilon)\,]\,h_m(\epsilon)\notag\\
&=F(-V_{lm})-F(\omega-V_{lm})~,
\label{41}
\end{align}
where
\be
F(\omega-V_{lm})=\frac{1}{2}\,\int\!d\epsilon\left[1-h_l(\epsilon+\omega)h_m(\epsilon)\right]
\label{Fomega}
\ee
is, for the integration with infinite limits, an even function of its argument and $V_{lm}=\mu_l-\mu_m$; specifically,
\be
F(\omega)=\omega\coth(\omega/2T)
\label{43}
\ee
for $h_l(\epsilon)$ from Eq.~(\ref{29}), which at $T=0$ becomes $F(\omega)=|\omega|$.

Substituting Eq.~(\ref{41}) in Eq.~(\ref{current}) and integrating over $x$ from 0+, as explained in the paragraph above Eq.~(\ref{41}), we have
\begin{align}
&\delta J_j=-\frac{e}{4\pi}\sum_{lm}\alpha_l\bigg\{\!-\!\pi B_{jlm}^{\prime\prime}\!\int\!d\epsilon\,f_m(\epsilon)\notag \\
&+\int_0^\infty\!\!d\omega\left[\,\chi^{\prime\prime}(\omega)F_{lm}^+(\omega)B_{jlm}^{\prime\prime}-\chi^\prime(\omega)F_{lm}^-(\omega)B_{jlm}^\prime\,\right]\bigg\}~,
\label{current-general}
\end{align}
where the functions $\chi^\prime(\omega)$ and $\chi^{\prime\prime}(\omega)$,
\begin{align}
\begin{bmatrix}
\chi^\prime(\omega) \\ \chi^{\prime\prime}(\omega)
\end{bmatrix}
&=
\begin{pmatrix}
{\rm Re} \\ {\rm Im}
\end{pmatrix}
\left\{\frac{2i}{v}\!\int_0^L\!\!dx\,e^{-2i\omega x/v}\right\}\notag\\
&=\frac{1}{\omega}
\begin{bmatrix}
1-\cos(2\omega L/v) \\ \sin(2\omega L/v)~
\end{bmatrix}~,
\label{45}
\end{align}
filter out the real ($B^{\prime}_{jlm}={\rm Re}\,B_{jlm}$) and imaginary ($B^{\prime\prime}_{jlm}={\rm Im}\,B_{jlm}$) parts of $B_{jlm}$, respectively, and
\be
F_{lm}^\pm(\omega)=F(\omega-V_{lm})\pm F(\omega+V_{lm})~.
\label{46}
\ee
The term $-\pi F(V_{lm})B_{jlm}^{\prime\prime}$ that would have been added to the integral in Eq.~(\ref{current-general}) if we had substituted  Eq.~(\ref{41}) in Eq.~(\ref{current}) and done the integration over $\omega$ and $x$ straightforwardly is absent---because of the exclusion of the point $x=0$ from the $x$ integration. Following from the same argument, since the integral in $\chi^{\prime\prime}(\omega)$ is defined as including the point $x=0$, the first term in the curly brackets in Eq.~(\ref{current-general}) compensates for the $\omega$ independent term in $F_{lm}^+(\omega)$.

Note that, by unitarity, the sum $\sum_m\!B^{\prime\prime}_{jlm}=0$, as follows directly from Eq.~(\ref{Ajlm}), which guarantees that the terms in $\delta J_j$ in Eq.~(\ref{current-general}) that are proportional to $B^{\prime\prime}_{jlm}$ vanish at equilibrium. The vanishing at equilibrium of the remaining part of $\delta J_j$ relies on $F_{lm}^-(\omega)$ being zero at equilibrium by construction.

\subsection{Y junction}
\label{s4b}

Equation (\ref{current-general}) gives the interaction-induced current for an arbitrary number of wires and an arbitrary form of the noninteracting $S$ matrix. We now apply Eq.~(\ref{current-general}) to the case of a Y junction with the noninteracting $S$ matrix obeying Eq.~(\ref{S33-matrix}), which is the model considered in Sec.~\ref{s3} within the picture of scattering off nonequilibrium Friedel oscillations.

The term in Eq.~(\ref{current-general}) that is proportional to $B_{jlm}^\prime$ represents a contribution to the current arising from the conventional renormalization of a junction \cite{Aristov2017} and will no longer be considered here. Our immediate goal, then, is to identify the chiral current $J_j^{\rm ch}$ by looking at the terms in Eq.~(\ref{current-general}) that are proportional to $B_{jlm}^{\prime\prime}$. Using the parametrization of the $S$ matrix from Eq.~(\ref{rr3tt3}), we see that
the nonzero components of $B_{jlm}^{\prime \prime }$ are all of the same modulus:
\be
B_{jlm}^{\prime\prime}=\mathcal{A}\,E_{jlm}~,
\label{47}
\ee
where $\cal A$ is given by Eq.~(\ref{caligra}) and the matrices $E_{jlm}$ are
\begin{align}
&E_{1lm}=\left(
\begin{array}{ccc}
0 & 0 & 0 \\
-1 & 0 & 1 \\
-1 & 1 & 0
\end{array}
\right)~,\quad
E_{2lm}=\left(
\begin{array}{ccc}
0 & -1 & 1 \\
0 & 0 & 0 \\
1 & -1 & 0
\end{array}
\right)~, \notag \\
&E_{3lm}=\left(
\begin{array}{ccc}
0 & 1 & -1 \\
1 & 0 & -1 \\
0 & 0 & 0
\end{array}
\right)~.
\label{48}
\end{align}
We recognize the first term in the curly brackets in Eq.~(\ref{current-general}) as associated with $\hat{G}^{\rm reg}$ [given by Eq.~(\ref{25}) for $T=0$].

The chiral current is then given by
\be
J_j^{\text{ch}}=-\frac{e}{4\pi}\sum_{lm}\alpha_lB_{jlm}^{\prime\prime}\!\int_0^\infty\!\!d\omega\,\chi^{\prime\prime}(\omega)F_{lm}^+(\omega)~.
\label{J-ch}
\ee
From Eq.~(\ref{J-ch}), the differential chiral conductance matrix $G_{jk}^{\rm ch}=e\,\partial J_j^{\rm ch}/\partial\mu_k$ is written as
\be
G_{jk}^{\rm ch}=-\frac{e^2}{4\pi}\sum_{lm}\alpha_lB_{jlm}^{\prime\prime}\!\int_0^\infty\!\!d\omega\,\chi^{\prime\prime}(\omega)\frac{\partial F_{lm}^+(\omega)}{\partial \mu_k}~.
\label{50}
\ee
For $T=0$, the integration over $\omega$ in Eq.~(\ref{50}) gives
\begin{align}
\int_0^\infty\!\!d\omega\,\chi^{\prime\prime}(\omega)\frac{\partial F_{lm}^+}{\partial\mu_k}&=2(\delta_{lk}-\delta_{mk})s_{lm} \notag\\
&\times\int_0^{|V_{lm}|}\!\!\frac{d\omega}{\omega}\sin\frac{2\omega L}{v}~,
\label{eq:saturInt}
\end{align}
where the sign function $s_{lm}$ is defined below Eq.~(\ref{26}), which in the limit of $|V_{lm}|L/v\to\infty$ reduces to
\be
\lim_{L\to\infty}\int_0^\infty\!\!d\omega\,\chi^{\prime\prime}(\omega)\frac{\partial F_{lm}^+}{\partial\mu_k}=\pi (\delta_{lk}-\delta_{mk})s_{lm}~.
\label{52}
\ee
The chiral conductance matrix thus takes the form
\be
G_{jk}^{\rm ch}=-\frac{1}{4}e^2\mathcal{A}\sum_{lm}\alpha_{l}E_{jlm}(\delta_{lk}-\delta_{mk})s_{lm}~,
\label{53}
\ee
which, upon inspection, coincides with $\hat{G}^{\rm ch}$ from Eq.~(\ref{26}). That is, the calculation we worked through in Secs.~\ref{s4a} and \ref{s4b} shows exactly how the physics of scattering off nonequilibrium Friedel oscillations (Sec.~\ref{s3}), which leads to the emergence of the chiral current, is encoded in the Keldysh formalism.

\subsection{Finite temperature}
\label{s4c}

We now turn to the case of finite $T$. As clearly seen from the structure of the expression for the pole-related current in Eq.~(\ref{21}), increasing $T$ leads to a suppression of the effect of nonequilibrium chirality. In the limit of $T/\Lambda\to 0$, the difference of the averages of the distribution functions in Eq.~(\ref{21}) obeys
\begin{align}
\langle f_jf_k\rangle-\langle f_jf_l\rangle=\frac{1}{2}\left(V_{kl}-V_{jk}\coth\frac{V_{jk}}{2T}+V_{jl}\coth\frac{V_{jl}}{2T}\right),\notag\\
\label{54a}
\end{align}
where the first term in the brackets on the right-hand side gives the $T$ independent linear conductance $\hat{G}^{\rm reg}$ from Eq.~(\ref{25}). For $|V_{jk}|,|V_{jl}|\ll T$, Eq.~(\ref{54a}) reduces to
\be
\langle f_jf_k\rangle-\langle f_jf_l\rangle\simeq \frac{1}{2}V_{kl}\left(1+\frac{V_{jk}+V_{jl}}{6T}\right)~,
\label{54}
\ee
where the second term in the brackets gives the leading term in the chiral conductance $\hat{G}^{\rm ch}$, which decreases as $1/T$ with increasing $T$ compared to Eq.~(\ref{26}). For the differential conductance, the expansion (\ref{54}) means the substitution
\be
s_{jk}\to V_{jk}/3T
\label{55}
\ee
in Eq.~(\ref{26}).

Within the Keldysh formulation, the generalization to finite $T$ proceeds with the use of Eq.~(\ref{43}) for $F(\omega)$ and the resulting change of $\partial F^+_{lm}/\partial\mu_k=2(\delta_{lk}-\delta_{mk})s_{lm}\Theta(|V_{lm}|-\omega)$ for zero $T$ [Eq.~(\ref{eq:saturInt})] to
\begin{align}
\frac{\partial F_{lm}^+}{\partial\mu_k}=(\delta_{lk}-\delta_{mk})\left[\,{\cal F}\left(\frac{\omega+V_{lm}}{2T}\right)-{\cal F}\left(\frac{\omega -V_{lm}}{2T}\right)\,\right]~,\notag\\
\label{dFdV}
\end{align}
where
\be
{\cal F}(x)=\coth x-x/\sinh^2x~.
\label{57}
\ee
Equation (\ref{52}) then changes to
\be
\lim_{L\to\infty}\int_0^\infty\!\!d\omega\,\chi^{\prime\prime}(\omega)\frac{\partial F_{lm}^+}{\partial\mu_k}=\pi(\delta_{lk}-\delta_{mk}){\cal F}\!\left(\frac{V_{lm}}{2T}\right)~,
\label{58}
\ee
so that $G_{jk}^{\rm ch}$ for arbitrary $T$ is given by Eq.~(\ref{53}) with the substitution of ${\cal F}(V_{lm}/2T)$ for $s_{lm}$. The asymptotic behavior of ${\cal F}(x)$ is: ${\cal F}(x)\to 2x/3$ for $|x|\ll 1$ and ${\cal F}(x)\to{\rm sgn}(x)$ for $|x|\gg 1$, which corresponds to Eq.~(\ref{55}) for the translation of the results for zero $T$ into those for large $T$. In particular, using Eqs.~(\ref{26}) and (\ref{55}), the expression for the chiral current (\ref{J3thetapsi}) changes to
\be
J_3\simeq -\frac{1}{8}e\alpha\,{\cal A}\frac{V^2}{T}
\label{59}
\ee
for $T\gg |V|$.

It is also worth noting that if $|V_{lm}|\ll T$ for a given pair of $l$ and $m$, but $T$ is much smaller than the bias voltage between either of two terminals $l,m$ and the remaining third terminal, the current distribution is essentially given by that for $T=0$ and the chemical potentials $\mu_l$ and $\mu_m$ assumed degenerate ($s_{lm}=0$). We will further comment on the finite-$T$ case---from the perspective of symmetry of $\hat{G}^{\rm ch}$---in Sec.~\ref{s5d}.

\section{Emergent chirality and the ``fundamental" conductance matrix}
\label{s5}

We now provide an additional way to quantify the phenomenon of emergent chirality by referring to the structure of the ``fundamental" conductance matrix mentioned in Sec.~\ref{s1}. As discussed at the end of Sec.~\ref{s3c}, the most general form of the $3\times 3$ conductance matrix for a triple junction [Eq.~(\ref{27})] is parametrized by three numbers plus a common multiplier, altogether four parameters, i.e., the rank of the matrix (\ref{27}) is two. The relation between the currents $J_{1,2,3}$ and the chemical potentials $\mu_{1,2,3}$ can thus be fully accounted for by means of a $2\times 2$ matrix. One of the useful formulations is based on the introduction of the linearly independent currents \cite{Aristov2017}
\begin{align}
&J_a=(J_1-J_2)/2~, \label{60}\\
&J_b=-J_3
\label{61}
\end{align}
and the differential conductance matrix
\be
\hat{\widetilde G}=\left(
\begin{array}{cc}
G_a & G_c+G_d \\
-G_c+G_d & G_b
\end{array}
\right)
\label{conductance-matrix}
\ee
which relates $J_{a,b}$ to the chemical potentials by
\begin{align}
&\!\!G_a=e\,\partial J_a/\partial V_a~,\quad G_b=e\,\partial J_b/\partial V_b~,\notag\\
&\!\!G_c+G_d=e\,\partial J_a/\partial V_b~,\quad -G_c+G_d=e\,\partial J_b/\partial V_a~,
\label{63}
\end{align}
where
\be
V_a=\mu_1-\mu_2,\quad V_b=\frac{1}{2}(\mu_1+\mu_2)-\mu_3~,
\label{64}
\ee
and the chemical potentials counted from their average value $\bar\mu$ are $\mu_1-\bar\mu=V_a/2+V_b/3$, $\mu_2-\bar\mu=-V_a/2+V_b/3$, and $\mu_3-\bar\mu=-2V_b/3$. The relation of $\hat{\widetilde G}$ to $\hat{G}$ is given by
\be
\hat{\widetilde{G}}=\frac{1}{4}
\begin{bmatrix}
G_{11}+G_{22}-G_{12}-G_{21}\, & \,2(G_{23}-G_{13}) \\
2(G_{32}-G_{31})\, & \,4G_{33}
\end{bmatrix}
\label{65}
\ee
with
\be
G_c=\frac{1}{2}(G_{12}-G_{21})~,\quad G_d=\frac{1}{2}(G_{11}-G_{22})~.
\label{66}
\ee
In Secs.~\ref{s5a}-\ref{s5c}, we focus on the case of $T=0$. The behavior of $G_c$ and $G_d$ at finite $T$ is discussed in Sec.~\ref{s5d}.

\subsection{Off-diagonal elements of $\hat{\widetilde G}$}
\label{s5a}

The significance of introducing $G_c$ and $G_d$ is that both of them are zero, so that the matrix $\hat{\widetilde{G}}$ is then diagonal, in the absence of interactions for the case of the $S$ matrix (\ref{S33-matrix}). A comparison of Eqs.~(\ref{27}) and (\ref{66}) shows that
\be
G_c\propto\xi_1-\xi_2~,\quad G_d\propto 1-\xi_3
\label{67}
\ee
describe time-reversal ($G_c$) and 1$\leftrightarrow$2 parity ($G_d$) symmetry breaking, as discussed below Eq.~(\ref{27}). Specifically, $G_c$ has the meaning of the interaction-induced nonlinear ``Hall" conductance, whereas $G_d$ quantifies the ``side diversion" current resulting from interaction-induced voltage-dependent inversion asymmetry between terminals 1 and 2. The current associated with inversion symmetry breaking can also be thought of in terms of the photogalvanic effect, viewed broadly, e.g., along the lines of Ref.~\cite{belinicher80}. Provided $G_c=G_d=0$ in the linear response (time-reversal and 1$\leftrightarrow$2 parity symmetric Hamiltonian), either or both of $G_c$ and $G_d$ being nonzero and chiral beyond the linear response is the essence of the phenomenon we called ``emergent chirality."

For the conductance matrix from Eq.~(\ref{26}), we have
\begin{align}
G_c=-\frac{1}{8}e^2\mathcal{A}\,&[\,\alpha_1(s_{12}+s_{31})+\alpha_2(s_{12}+s_{23}) \notag \\
&+\alpha_3(s_{23}+s_{31})\,]~,
\label{68}
\end{align}
The emergence of $G_c\neq 0$ is a truly nonequilibrium phenomenon, with all terms in Eq.~(\ref{68}) depending on the signs of the voltages, for arbitrary $\alpha_{1,2,3}$. By contrast, if the interacting part of the Hamiltonian is not 1$\leftrightarrow$2 parity symmetric ($\alpha_1\neq\alpha_2$), then $G_d$ is a sum of both chiral and nonchiral terms, $G_d=G_d^{\rm ch}+G_d^{\rm lin}$, where the nonchiral term $G_d^{\rm lin}$ exists already in the linear response (i.e., is not dependent on the signs of any voltages). Both the scale-dependent and pole contributions to $G_d^{\rm lin}$, proportional to $B_{jlm}^\prime$ and $B_{jlm}^{\prime\prime}$ from Eq.~(\ref{current-general}), respectively, are nonzero for $\alpha_1\neq\alpha_2$. From Eq.~(\ref{26}), the chiral term $G_d^{\rm ch}$ reads
\be
G_d^{\rm ch}=\frac{1}{8}e^2\mathcal{A}\,[\,(\alpha_1+\alpha_2)s_{12}-\alpha_3(s_{23}+s_{31})\,]~.
\label{69}
\ee

For $\mu_3$ lying between $\mu_1$ and $\mu_2$, we have $s_{12}=s_{13}=-s_{23}={\rm sgn}V_a$, which gives
\begin{align}
&G_c=\frac{1}{4}e^2{\cal A}\,\alpha_3\,{\rm sgn}V_a~,
\label{Hall-conductance}\\
&G_d^{\rm ch}=\frac{1}{8}e^2{\cal A}\,(\alpha_1+\alpha_2+2\alpha_3)\,{\rm sgn}V_a~.
\label{71}
\end{align}
For $V_b=0$ (bias $\mu_1-\mu_2$ applied symmetrically with respect to $\mu_3$), $\alpha_1=\alpha_2=\alpha$, and $\alpha_3=0$, which is the voltage setup and the choice of $\alpha_{1,2,3}$ considered in Secs.~\ref{s3a} and \ref{s3b}, the expression for $J_3$ in terms of the elements of the matrix (\ref{conductance-matrix}) becomes (with $G_d$ given entirely by $G_d^{\rm ch}$)
\be
J_3=-\frac{1}{e}\,G_dV_a=-\frac{1}{4}e\alpha\,{\cal A}|V_a|~,
\label{72}
\ee
which coincides with Eq.~(\ref{J3thetapsi}). Equation (\ref{72}) thus tells us that the emergence of the chiral current in the symmetrically biased tunneling-tip setup with $\alpha_3\to 0$ is the effect of broken 1$\leftrightarrow$2 parity symmetry, controlled by the conductance $G_d^{\rm ch}$. The same conclusion can also be drawn from the chiral part of the $3\times 3$ conductance matrix [Eq.~(\ref{26})], which is then characterized, in terms of the parameters $\xi_{1,2,3}$ from Eq.~(\ref{27}), by $\xi_1=\xi_2=0$ and $\xi_3=-1$.

As was already noted below Eq.~(\ref{22}), the current $J_3$ does not depend on $\alpha_3$ for an arbitrary distribution of voltages. For the case of $\mu_3$ between $\mu_1$ and $\mu_2$ [Eqs.~(\ref{Hall-conductance}) and (\ref{71})], this shows up in the cancellation of the $\alpha_3$ dependent terms in the combination $G_d^{\rm ch}-G_c$, which is probed in this type of measurement. In particular, for $V_b=0$ and $\alpha_1=\alpha_2=\alpha$, the current $J_3$ is given by the last expression in Eq.~(\ref{72}) for arbitrary $\alpha_3$. For $\alpha_3\neq 0$, the emergence of nonzero $J_3$ for $V_b=0$ is thus a combined effect of nonequilibrium 1$\leftrightarrow$2 parity and time-reversal symmetry breaking, in which the role of time-reversal symmetry breaking is to exactly cancel the $\alpha_3$ contribution to $J_3$. This example demonstrates the inherent relationship between the two types of symmetry, formalizable as combined parity-time symmetry in the chiral current.

It is worth noting that the emergence of off-diagonal elements of the matrix $\hat{\widetilde G}$ is generic for arbitary $\mu_{1,2,3}$ except for the special case of $\mu_1=\mu_2$. Specifically, if $\mu_1=\mu_2$, then $G_d^{\rm ch}=0$ identically for arbitrary $\alpha_{1,2,3}$. If, additionally, $\alpha_1=\alpha_2$, then also $G_c=0$ at $\mu_1=\mu_2$. By permutation of the wire indices, the vanishing of $G_c$ occurs for two arbitrary chemical potentials $\mu_i$ and $\mu_j$ being degenerate if $\alpha_i=\alpha_j$.

It is also worthwhile to comment on the difference between $G_c$ and $G_d$ regarding their dependence on $\alpha_3$ in the tunneling-tip setup ($\alpha_1=\alpha_2$). As follows from Eq.~(\ref{68}), $G_c$ for any given distribution of voltages with nondegenerate chemical potentials depends on only one out of three interaction constants $\alpha_{1,2,3}$, namely the one in the wire with the ``intermediate" chemical potential (which lies between the largest and lowest ones). Specifically, $G_c\propto\alpha_3$ for $\mu_3$ between $\mu_1$ and $\mu_2$ [Eq.~(\ref{Hall-conductance})] and $G_c\propto\alpha$ in the case of $\alpha_1=\alpha_2=\alpha$ for any other distribution of voltages (with the exception of two chemical potentials being degenerate---then $G_c$ is given by a half-sum of the interaction constants in two wires with the degenerate chemical potentials). This is why it is possible to arrange the voltages to produce the chiral current in Eq.~(\ref{72}) that specifically probes $G_d$, with no admixture of $G_c$, in the limit of $\alpha_3\to 0$. By contrast, $G_d$ for $\alpha_1=\alpha_2$ does not vanish in the limit of $\alpha_3\to 0$ for any ``nondegenerate" distribution of $\mu_{1,2,3}$.

In fact, the vanishing of $G_d^{\rm ch}$ while $G_c\neq 0$ requires, for a distribution of voltages with nondegenerate chemical potentials, that interactions of different signs be present in the system. Specifically, in the nondegenerate situation, $G_d^{\rm ch}=0$ for: either (i) $\alpha_1=-\alpha_2$ and $\alpha_3(s_{23}+s_{31})=0$, where the latter condition means $\alpha_3=0$ for an arbitrary distribution of voltages or arbitrary $\alpha_3$ for $\mu_3$ lying above or below both $\mu_1$ and $\mu_2$, or (ii) $\alpha_1+\alpha_2=-2\alpha_3$ and $\mu_3$ lying between $\mu_1$ and $\mu_2$. Therefore, to separate out the effect of emergent chirality that is due to nonequilibrium time-reversal symmetry breaking without fine-tuning the Hamiltonian with regard to the interaction constants of opposite signs, one has to perform at least two current measurements with different arrangements of voltages and compare the results, as we explain below.

\subsection{Properties of $\hat{\widetilde G}$ with respect to flipping the signs of voltages}
\label{s5b}

Since the conductance matrix $\hat{\widetilde G}$ is a function of voltages, the current measurements for different voltage distributions yield, generically, different sets of the nonlinear conductances. One consequence is that the symmetry properties of $\hat{\widetilde G}$ include symmetry with respect to flipping the sign of the bias voltage between two terminals. A question arises, then, if the chiral conductances $G_c^{\rm ch}$ and $G_d^{\rm ch}$ are separately measurable by only manipulating the signs of the voltages. Adding up, for each of the wires, the currents before and after a simultaneous change of the signs of $V_a$ and $V_b$ filters out the chiral components of $J_a$ and $J_b$. Generically, however, these are functions of four chiral conductances $G_{a,b,c,d}^{\rm ch}$---because the nonlinearity is present in both the diagonal and nondiagonal elements of $\hat{\widetilde G}$. It follows that, for a generic distribution of voltages, this procedure does not yield $G_c^{\rm ch}$ and $G_d^{\rm ch}$.

At this point, it is instructive to look at a simple example in which the junction is pushed out of equilibrium by adding voltage $V$ at only one of three terminals. Assume also that $\alpha_1=\alpha_2=\alpha$. As mentioned above, for $\mu_1=\mu_2$ both $G_d$ and $G_c$ are zero in that case. If it is $\mu_3$ that remains degenerate with either $\mu_1$ or $\mu_2$, then changing $V\to -V$ allows one to measure two combinations of four chiral conductances, namely $G_a^{\rm ch}-G_c^{\rm ch}/2-G_d^{\rm ch}/2$ and $G_b^{\rm ch}/2+G_c^{\rm ch}-G_d^{\rm ch}$, by measuring the chiral components of $J_a$ and $J_b$, respectively. The effect of $G_a^{\rm ch}$ and $G_b^{\rm ch}$ is seen to intertwine with that of $G_c^{\rm ch}$ and $G_d^{\rm ch}$. In this example, both $G_a^{\rm ch}$ and $G_b^{\rm ch}$ are nonzero, also for $\alpha_3=0$:
\begin{align}
&G_a^{\rm ch}=\frac{1}{16}e^2{\cal A}\,(\alpha+2\alpha_3)\,{\rm sgn}V_b~, \notag\\
&G_b^{\rm ch}=-\frac{1}{4}e^2{\cal A}\,\alpha\,{\rm sgn}V_b
\label{73}
\end{align}
for the voltage $V=V_b/2$ applied to either terminal 1 or terminal 2.

The invariance of $G_{a,b}^{\rm ch}$ with respect to exchanging $\mu_1$ and $\mu_2$ in Eqs.~(\ref{73}) is a particular example of general (for $\alpha_1=\alpha_2$) symmetry of $\hat{\widetilde G}$ as a function of $V_{a,b}$, as follows from Eq.~(\ref{26}):
\begin{align}
&G_{a,b}^{\rm ch}\left(-V_a,V_b\right)=G_{a,b}^{\rm ch}\left(V_a,V_b\right)~,\notag\\
&G_{c,d}^{\rm ch}\left(-V_a,V_b\right)=-G_{c,d}^{\rm ch}\left(V_a,V_b\right)~,
\label{74}
\end{align}
which translates into the relation between the currents:
\begin{align}
&J_a^{\rm ch}\left(-V_a,V_b\right)=-J_a^{\rm ch}\left(V_a,V_b\right)~,\notag\\
&J_b^{\rm ch}\left(-V_a,V_b\right)=J_b^{\rm ch}\left(V_a,V_b\right)~.
\label{75}
\end{align}
Note the existence of the ``cross-term" (which depends on both voltages $V_a$ and $V_b$) in $J_a^{\rm ch}$, with
\be
\left(G_c^{\rm ch}+G_d^{\rm ch}\right)V_b\propto V_b\,{\rm sgn}V_a~.
\label{76}
\ee
Note also that $J_a^{\rm ch}$ changes sign with exchanging $\mu_1$ and $\mu_2$ ($V_a\to -V_a$ with $V_b$ held fixed), which makes it indistinguishable from the noninteracting contribution to $J_a$ under this symmetry operation.

\subsection{Measurement protocol to extract the chiral components of $G_c$ and $G_d$}
\label{s5c}

Having described the general behavior of $\hat{\widetilde G}$ as a function of $V_{a,b}$ [Eqs.~(\ref{74})], exemplified by the case of $V_b=\pm V_a/2$ [Eqs.~(\ref{73})], let us turn to another special case, in which no degeneracy in the chemical potentials is left. Let $\mu_3$ be squeezed between $\mu_1$ and $\mu_2$ ($|V_b|<|V_a|/2$). As follows from Eq.~(\ref{26}), the diagonal chiral conductances $G_a$ and $G_b$ are both zero for this arrangement of voltages for the tunneling tip setup with $\alpha_1=\alpha_2$ and arbitrary $\alpha_3$. That is,
\begin{align}
&J_a^{\rm ch}=(G_c+G_d)V_b~,
&J_b^{\rm ch}=(-G_c+G_d)V_a~,
\label{77}
\end{align}
with $G_c$ and $G_d$ given by Eqs.~(\ref{Hall-conductance}) and (\ref{71}), both proportional to ${\rm sgn}V_a$. The vanishing of $G_{a,b}$ for $|V_b|<|V_a|/2$ makes a big difference compared to their nonzero values for $|V_b|=|V_a|/2$ in Eq.~(\ref{73}). For $V_b=0$ [the case describable by Eqs.~(\ref{77})], we return to the symmetric setup in which the existence of a nonzero current $J_b$ is perhaps the most remarkable manifestation of the phenomenon of emergent chirality.

For $V_b\neq 0$, with $V_b$ parametrizing the difference of the spacings separating $\mu_3$ from $\mu_1$ and $\mu_2$, both $G_c$ and $G_d$ are in play---and, in Eqs.~(\ref{77}), only these two. It follows that if both $J_a^{\rm ch}$ and $J_b^{\rm ch}$ are known for given $V_a$ and $V_b$, then $G_c$ and $G_d$ can be determined separately from Eqs.~(\ref{77}). An important point, following from the relations (\ref{75}) is that, to extract the chiral components of both $J_a$ and $J_b$, one should flip the signs of all voltages---exchanging $\mu_1$ and $\mu_2$ does not suffice, as explained below Eq.~(\ref{76}). The condition for this ``protocol" being useful for the determination of $G_c^{\rm ch}$ and $G_d^{\rm ch}$ is the placement of $\mu_3$ between $\mu_1$ and $\mu_2$ [excluding the end points of the interval, as demonstrated in Eqs.~(\ref{73}) by the emergence of nonzero $G_a^{\rm ch}$ and $G_b^{\rm ch}$ at these points when $\mu_3$ varies with respect to $\mu_1$ and $\mu_2$, with the starting point between the two].

The procedure of determining $G_c^{\rm ch}$ and $G_d^{\rm ch}$ by relying on Eqs.~(\ref{77}) elucidates the meaning of these conductances and is essentially equivalent to the measurement based on the direct definition of $G_c$ and $G_d$ in Eqs.~(\ref{66}) in terms of the partial derivatives $\partial J_j/\partial\mu_k$. For example, for the conductance matrix (\ref{26}), which only changes in a stepwise manner with varying $\mu_{1,2,3}$ when two chemical potentials ``cross" each other, a discretized version of the differentiation relates the emergence of nonzero $G_c$ at nonequilibrium to the inequality
\begin{align}
&J_1(\mu_1,\mu_2+\delta V,\mu_3)-J_2(\mu_1+\delta V,\mu_2,\mu_3)\notag\\
&\neq J_1(\mu_1,\mu_2,\mu_3)-J_2(\mu_1,\mu_2,\mu_3)
\label{78}
\end{align}
for $\delta V$ the addition of which does not change the mutual order of $\mu_{1,2,3}$. Similarly for $G_d$:
\begin{align}
&J_1(\mu_1+\delta V,\mu_2,\mu_3)-J_2(\mu_1,\mu_2+\delta V,\mu_3)\notag\\
&\neq J_1(\mu_1,\mu_2,\mu_3)-J_2(\mu_1,\mu_2,\mu_3)~.
\label{79}
\end{align}
Both inequalities become equalities in the noninteracting limit for the time-reversal and 1$\leftrightarrow$2 parity symmetric $S$ matrix (\ref{S33-matrix}). The measurement protocol to determine $G_c$ and/or $G_d$ thus generally (not implying that $G_c$ and $G_d$ are chiral) includes measuring the currents for three different arrangements of the chemical potentials: $(\mu_1,\mu_2,\mu_3)$, $(\mu_1+\delta V,\mu_2,\mu_3)$, and $(\mu_1,\mu_2+\delta V,\mu_3)$. For the case of $\mu_3$ between $\mu_1$ and $\mu_2$ and the relation (\ref{77}), the protocol reduces to two different sets of the voltages: $(V_a,V_b)$ and $(-V_a,-V_b)$.

\subsection{Temperature dependence of $G_c$ and $G_d$}
\label{s5d}

We now demonstrate that the dependence of the nondiagonal conductances $G_c$ and $G_d$ on $\alpha_{1,2,3}$ and the voltages changes in an essential way with increasing $T$. For this purpose, we return to the case of large $T$, considered (together with the general case of arbitrary $T$) in Sec.~\ref{s4c}, by calculating $G_c$ and $G_d$ to order ${\cal O}(1/T)$. Substituting Eq.~(\ref{55}) in Eqs.~(\ref{68}) and (\ref{69}), we have for $T\gg |V_a|,|V_b|$:
\begin{align}
&G_c\simeq\frac{e^2{\cal A}}{24\,T}\left(\alpha_1V_{23}+\alpha_2V_{31}+\alpha_3V_{12}\right)~,
\label{80}\\
&G_d\simeq\frac{e^2{\cal A}}{24\,T}\left(\alpha_1+\alpha_2+\alpha_3\right)V_{12}~.
\label{81}
\end{align}
Note that in the tunneling-tip setup with $\alpha_1=\alpha_2=\alpha$ and $\alpha_3\neq\alpha$ both $G_c$ and $G_d$ in the large-$T$ limit only depend on $V_a$, independently of the mutual position of $\mu_{1,2,3}$:
\begin{align}
&G_c\simeq-\frac{e^2{\cal A}}{24\,T}\left(\alpha-\alpha_3\right)V_a~,
\label{82}\\
&G_d\simeq\frac{e^2{\cal A}}{24\,T}\left(2\alpha+\alpha_3\right)V_a~.
\label{83}
\end{align}
This is in contrast to the zero-$T$ limit, where $G_c$ and $G_d$ generically depend on both $V_a$ and $V_b$, except for the case of $\mu_3$ lying between $\mu_1$ and $\mu_2$ [Eqs.~(\ref{Hall-conductance}) and (\ref{71})]. If $\alpha_3=\alpha$, the expansion of $G_c$ in powers of $1/T$ starts at order $V_{12}V_{23}V_{31}/T^3$.

From Eqs.~(\ref{82}) and (\ref{83}), substituted in Eq.~(\ref{conductance-matrix}), we recover Eq.~(\ref{59}) for $J_3$ in the symmetrically biased junction with $V_b=0$. Note that, as a manifestation of the general rule formulated below Eq.~(\ref{22}), the terms proportional to $\alpha_3$ cancel out in $J_3$, independently of whether $T$ is zero or not. Note also that, for nonzero $T$, time-reversal symmetry is generically broken $(G_c\neq 0)$ in this setup, even for $\alpha_3=0$, but $G_c$ vanishes in the limit of $T\to 0$ if $\alpha_3=0$ [Eq.~(\ref{Hall-conductance})].

\section{Higher-order renormalization}
\label{s6}

We now briefly discuss the interaction-induced renormalization of the chiral current (\ref{J3thetapsi}) [or (\ref{72}) for that matter] at zero $T$ in the vicinity of the stable critical point at which all three wires are decoupled from each other (``point $N$"). The global behavior of the renormalization-group flow to this point at nonequilibrium was discussed in the limit of weak interaction in Refs.~\cite{Aristov2017,Aristov2018}, with the flow being stopped by nonequilibrium in an intricate way, distinctly different from the effect of temperature. We also provide a similar result for scaling near the unstable fixed point where wire 3 is decoupled from the ballistic 1+2 wire (``point $A$").

We can infer how the factor $\cal A$, which quantifies the amplitude of $J_3$ in Eq.~(\ref{J3thetapsi}), is renormalized by first writing down $G_a$ and $G_b$ in the absence of interaction in terms of the angles $\theta$ and $\psi$ [by using $G_{ij}=(e^2/2\pi)(\delta_{ij}-|S_{ij}|^2)$ and Eq.~(\ref{65})]:
\be
G_a=\frac{e^2}{4\pi}(1-\cos\theta\cos\psi)~,\quad G_b=\frac{e^2}{2\pi}\sin^2\theta~.
\label{84}
\ee
From Eq.~(\ref{caligra}), the bare value of $\cal A$ is then expressible in terms of the dimensionless conductances $\bar{G}_{a,b}=2\pi G_{a,b}/e^2$ as
\be
\mathcal{A}=\frac{s}{4}\,\bar{G}_b\left[\,4\bar{G}_a\left(1-\bar{G}_a\right)-\bar{G}_b\,\right]^{1/2}~,
\label{85}
\ee
where $s={\rm sgn}(\cos\theta\sin\psi)$. At both fixed points $N$ and $A$, ${\cal A}=0$ [recall also the comment below Eq.~(\ref{J3thetapsi})].

In the presence of interaction, the renormalization-group flow under nonequilibrium conditions generically breaks time-reversal and 1$\leftrightarrow$2 symmetry of the conductance matrix (but does not, by itself, lead to the emergence of chiral currents, as was already mentioned in Sec.~\ref{s1}). As a result, the renormalized $S$ matrix (even without the pole-related terms) is generically not parametrized as in Eq.~(\ref{rr3tt3}) \cite{Aristov2017}. However, the renormalization preserves both symmetries for $V_b=0$, as can be seen from Eq.~(14) of Ref.~\cite{Aristov2017}. Therefore, for the main wire biased symmetrically with respect to the tip wire, Eq.~(\ref{85}) gives the relation between the running values of $\cal A$ and $\bar{G}_{a,b}$ at each point of the renormalization-group flow. With the renormalization included, the chiral current $J_3$ from Eq.~(\ref{72}) is then representable, for $T=0$, as
\be
J_3=-\frac{s}{16}\,e\alpha\,\bar{G}_b\!\left[\,4\bar{G}_a\left(1-\bar{G}_a\right)-\bar{G}_b\,\right]^{1/2}|V_a|~,
\label{86}
\ee
where $\bar{G}_{a,b}$ are understood as fully renormalized.

In the neighborhood of the stable point $N$ (where both $\bar{G}_{a,b}\ll 1$), using the results of Ref.~\cite{Aristov2017} for $\bar{G}_b$ and $4\bar{G}_a-\bar{G}_b$,
\be
\bar{G}_b\propto |V_a|^{\alpha+\alpha_3},\quad 4\bar{G}_a-\bar{G}_b\propto |V_a|^{2\alpha}~,
\label{87}
\ee
we find from Eq.~(\ref{86}):
\be
J_3\propto |V_a|^{1+2\alpha+\alpha_3}~.
\label{88}
\ee
Similarly, the renormalization-group flow near the unstable point $A$ (where $1-\bar{G}_a\ll 1$ and $\bar{G}_b\ll 1$) obeys \cite{Aristov2017}
\be
1-\bar{G}_a\propto |V_a|^{-2\alpha}~,\quad \bar{G}_b\propto |V_a|^{\alpha_3}
\label{89}
\ee
for $1-\bar{G}_a\gg\bar{G}_b$ in the ``runaway" domain of the flow \cite{remark1}, and we obtain
\be
J_3\propto |V_a|^{1-\alpha+\alpha_3}~.
\label{90}
\ee
As follows from Eq.~(\ref{88}), the conductance $e\,\partial J_3/\partial V_a=G_c-G_d$ [for $V_b=0$, as assumed in Eq.~(\ref{86})] as a function of $V_a$ changes sign at $V_a=0$, with the steplike jump being ``smoothed" by the renormalization for the case of repulsive interaction with $2\alpha+\alpha_3>0$.

\section{Effective chiral model}
\label{s7}

As mentioned in Sec.~\ref{s1}, the phenomenon of emergent chirality is reminiscent of the transport properties of a junction in the absence of interaction but in the presence of a magnetic flux and/or built-in asymmetry between wires 1 and 2. We now compare the interaction-induced chiral conductances $G_c$ and $G_d$ for the time-reversal and 1$\leftrightarrow$2 symmetric bare $S$ matrix---namely we take as an example those from Eqs.~(\ref{Hall-conductance}) and (\ref{71})---with their counterparts for a noninteracting junction with these symmetries broken ``by construction."

\begin{figure}[t]
\includegraphics[width=0.9\columnwidth]{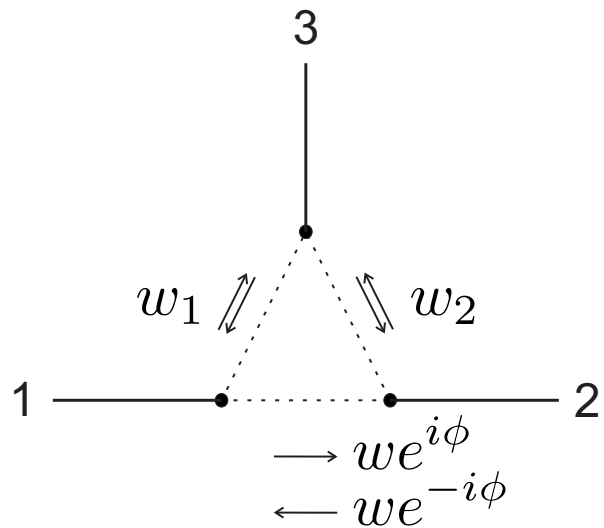}
\caption{Triple junction in the noninteracting model with a magnetic flux and 1$\leftrightarrow$2 parity asymmetry to mimic the effect of emergent chirality induced by interactions at nonequilibrium in the model with the time-reversal and 1$\leftrightarrow$2 parity symmetric Hamiltonian. The hopping amplitudes that couple the end points of wires 1,2,3 are marked together with the direction of hopping.}
\label{f3}
\end{figure}

For concreteness, we use the model of a Y junction consisting of three end points of wires 1,2,3, with these points connected by hopping matrix elements (Fig.~\ref{f3}). The $S$ matrix of the junction can be parametrized as
\be
\hat{S}=(1-i\hat{W})^{-1}(1+i\hat{W})~,
\label{91}
\ee
where the matrix $\hat W$ of dimensionless hopping amplitudes is given by
\be
\hat{W}=
\begin{pmatrix}
0 & we^{-i\phi} & w_1 \\
we^{i\phi} & 0 & w_2 \\
w_1 & w_2 & 0
\end{pmatrix}
\label{92}
\ee
with real numbers $w,w_1,w_2$. The points connected by hopping are vertices of a triangle threaded by the magnetic flux $\phi$ in units of the flux quantum $hc/e$ (restoring here $\hbar=h/2\pi$). If $\phi\neq 0$ (modulo $\pi$), scattering at the junction is not time-reversal symmetric; if $w_1\neq w_2$, it is not 1$\leftrightarrow$2 parity symmetric. In the case of identical links with $w=w_1=w_2$ and nonzero $\phi$, this model was introduced in Ref.~\cite{Oshikawa2006} for studying the role of $\phi$ in the interaction-induced renormalization of the junction. It is perhaps worth noting once again that, by contrast, our model in the noninteracting limit [Eq.~(\ref{S33-matrix})] is time-reversal symmetric, so that the effective magnetic flux to compare with $\phi$ in Eq.~(\ref{92}) is solely induced by interactions under nonequilibrium conditions \cite{remark3,buccheri18}.

Consider first the case of $\phi\neq 0$ and $w_2=w_1$. The fundamental conductance matrix (\ref{conductance-matrix}) for the noninteracting model $\hat{\widetilde G^\prime}$ (marked by the prime sign, together with its elements) is then antisymmetric with $G_c^\prime\neq 0$ and $G_d^\prime=0$. Since the chiral current $J_3$ in Eq.~(\ref{86}) is proportional to $G_b$, it is useful to write down the relation between $G_c^\prime$ and $G_b^\prime$, which is
\be
G_c^\prime=\frac{w\sin\phi}{1+w^2}\,G_b^\prime~.
\label{93}
\ee
This should be compared with $G_c$ from Eq.~(\ref{Hall-conductance}), which is representable by means of Eq.~(\ref{85}) as
\be
G_c=\frac{s}{16}\,e^2\alpha_3\bar{G}_b\left[\,4\bar{G}_a\left(1-\bar{G}_a\right)-\bar{G}_b\,\right]^{1/2}{\rm sgn}V_a~.
\label{94}
\ee
To simplify the comparison, let us look at the relation between $G_c$ and $G_c^\prime$ near the two fixed points (stable $N$ and unstable $A$) discussed in Sec.~\ref{s6}.

In the vicinity of point $N$, both $|t_3|,|t|\ll 1$, i.e., $\bar{G}_a=|t|^2+|t_3|^2/2\ll 1$ and $\bar{G}_b=2|t_3|^2\ll 1$, so that the difference of two terms in the square brackets of Eq.~(\ref{94}) reduces to $4|t|^2$, with no competition between $|t|$ and $|t_3|$. Similarly, in the noninteracting model, $|t|,|t_3|\ll 1$ means $|w|,|w_1|\ll 1$ with $|w|\simeq |t|/2$ and $|w_1|\simeq |t_3|/2$. As a result, $G_c$ and $G_c^\prime$ have a similar structure near point $N$, both proportional to $|t||t_3|^2$, with $\sin\phi$ independent of the amplitude of $t$ or $t_3$, namely
\be
\sin\phi\to\frac{\pi}{2}\alpha_3{\sf s}~,
\label{95}
\ee
where ${\sf s}={\rm sgn V}_{\! a}\times s\times{\rm sgn}\,w$ is a product of three sign functions [with $s$ defined below Eq.~(\ref{85})]. The effective magnetic flux is thus given by the interaction strength in wire 3 and its sign changes with flipping the sign of the voltage between terminals 1 and 2. It is worth recalling that the calculation in Secs.~\ref{s3} and \ref{s4} was done to first order in $\alpha_3$, so that Eq.~(\ref{95}) only establishes a linear relation between $\phi$ and $\alpha_3$ for $|\alpha_3|\ll 1$, which is sufficient for our purposes here \cite{remark2}.

Near point $A$, where $1-|t|\ll 1$ and $|t_3|\ll 1$, in the runaway region of the renormalization-group flow, we have $1\gg 1-\bar{G}_a\simeq 2(1-|t|)\gg\bar{G}_b=2|t_3|^2$. The expression in the square brackets of Eq.~(\ref{94}) is then given by $8(1-|t|)$. The scattering amplitudes of the noninteracting model obey, for $|t_3|^2\ll 1-|t|\ll 1$, the relation $(1-|w|)^2\simeq 2(1-|t|)$ and $|w_1|\simeq |t_3|/\sqrt{2}$. Following the same scheme of relating $G_c$ and $G_c^\prime$ as for point $N$, we obtain
\be
\sin\phi\to\frac{\pi}{\sqrt{2}}\alpha_3(1-|t|)^{1/2}{\sf s}~.
\label{96}
\ee
The structure of the relation between $G_c$ and $G_c^\prime$ in the vicinity of point $A$ is seen to be different compared to point $N$, namely the effective flux in Eq.~(\ref{96}) depends on the distance to the fixed point.

Turn now to the case of $\phi=0$ and $w_1\neq w_2$. The matrix $\hat{\widetilde G^\prime}$ is then symmetric, i.e., $G_c^\prime=0$, but $G_d^\prime\neq 0$. The analogue of Eq.~(\ref{93}), now for the relation between $G_d^\prime$ and $G_b^\prime$, reads
\be
G_d^\prime=-\frac{1}{2}\frac{(w_1^2-w_2^2)(1-w^2)}{(w_1^2+w_2^2)(1+w^2)}\,G_b^\prime~.
\label{97}
\ee
Referring to Eqs.~(\ref{71}) and (\ref{85}), we have (for $\alpha_1=\alpha_2=\alpha$):
\be
G_d=\frac{s}{16}\,e^2(\alpha+\alpha_3)\bar{G}_b\left[\,4\bar{G}_a\left(1-\bar{G}_a\right)-\bar{G}_b\,\right]^{1/2}{\rm sgn}V_a~.
\label{98}
\ee
Repeating the sequence of steps that led to Eqs.~(\ref{95}) and (\ref{96}), the effective anisotropy parameter $(w_1-w_2)/(w_1+w_2)$ near either point $N$ or point $A$ (in the runaway domain) is related to $\alpha$ and $\alpha_3$ by
\be
\frac{w_1-w_2}{w_1+w_2}\to -\frac{s\pi}{4}(\alpha+\alpha_3)|t|\,{\rm sgn}V_a~,
\label{99}
\ee
with $|t|\ll 1$ in the former case and $|t|=1$ in the latter. Note that the effective flux is nonzero at point $N$ ($|t|\to 0$) and vanishes at point $A$ ($|t|\to 1$), whereas for the effective anisotropy parameter the situation is the opposite: it vanishes at point $N$ and is nonzero at point $A$.

It is worthwhile to mention that the matrix $\hat{W}$ of the form
\be
\hat{W}=
\begin{pmatrix}
0 & we^{-i\phi} & w_1e^{-i\phi/2} \\
we^{i\phi} & 0 & w_1e^{i\phi/2} \\
w_1e^{i\phi/2} & w_1e^{-i\phi/2} & 0
\end{pmatrix}~,
\label{100}
\ee
which corresponds to a zero magnetic flux through the triangle junction, breaks (for $\phi\neq 0$ modulo $2\pi$) both time-reversal and 1$\leftrightarrow$2 parity symmetries of the $S$ matrix, but does not break these in the conductance matrix $\hat{\widetilde G^\prime}$, in which both $G_c^\prime$ and $G_d^\prime$ are zero. It is thus a nonzero effective magnetic flux that is inherently related to the emergence of finite $G_c$ in the interacting problem at nonequilibrium (which justifies naming $G_c$ the ``Hall conductance")---an inhomogeneous effective magnetic field with zero mean inside the junction does not suffice. We reiterate that it is the property of the Hall conductance $G_c$ being an odd function of the voltage, as opposed to the equilibrium chiral model \cite{Oshikawa2006}, that leads to the unidirectionality of the chiral currents in the leads, which constitutes the essence of emergent chirality introduced in our paper.

To give a finishing touch to the comparison to the chiral model \cite{Oshikawa2006}, it is also worth mentioning that the chiral fixed point from Ref.~\cite{Oshikawa2006}, at which the incoming currents are fully diverted in either the clockwise or counterclockwise direction, cannot be realized in a symmetric junction with $\phi=0$ by manipulating the voltages (Sec.~\ref{s5c}). Indeed, in terms of the fundamental conductances, the chiral fixed point corresponds to $G_a=3/4$, $G_b=1$, $G_c=\pm 1/2$, $G_d=0$ in units of $e^2/2\pi$. As can be seen from Eqs.~(\ref{68}) and (\ref{69}), for the symmetric junction with $\alpha_1=\alpha_2=\alpha_3$, if $G_c\neq 0$, then necessarily $G_d\neq 0$, i.e., the junction becomes asymmetric. It is important here that the nonequilibrium-induced $G_c$ and $G_d$, despite being scale-independent, are not corrections to the ultraviolet (bare) values of $G_c=G_d=0$, i.e., a finite $G_c$ in our problem does not bring the renormalization-group flow into the basin of the chiral fixed point. As described in Sec.~\ref{s6}, the renormalization-group flow is the same as at equilibrium (no effective flux affecting the flow), with $G_c$ and $G_d$ being ``infrared quantities" which only appear after the renormalization is done.

\section{Conclusion}
\label{s8}

We have presented a theory of the phenomenon we named ``emergent chirality," a distilled example of which is the emergence of a nonzero current in the ``side-wire" (tunneling tip wire in the electron tunneling experiment) driven by a current in the symmetrically biased main wire, as illustrated in Fig.~\ref{f1}. This result is quite remarkable as the current in the side-wire does not depend on the sign of the voltage and is exactly zero in the linear response, i.e., this is an essentially nonequilibrium phenomenon which breaks time-reversal and/or parity symmetry that exists at the level of the Hamiltonian of the system.

In the picture we have developed, the chirality of the current is inherently linked to the presence of electron-electron interactions. We have given a detailed discussion of emergent chirality from the perspective of electron scattering off nonequilibrium Friedel oscillations and also performed a formal perturbation theory calculation in the Keldysh technique. Perhaps one of the most important points to emphasize is that this phenomenon is totally different from the conventional interaction-induced renormalization, which, in particular, gives exactly zero current in the side-wire in Fig.~\ref{f1}. Rather, as opposed to virtual processes responsible for the renormalization, it is entirely due to real processes, one of the conceptually important peculiarities of which is that they are not inelastic scattering and give rise to the chiral current already at first order in the interaction strength.

Before concluding the paper, it is interesting to mention that the emergent chirality is to an extent phenomenologically similar to Bernoulli's effect that occurs in a Y junction of fluid-filled pipes, where pumping the fluid through the ``main" pipe forces the fluid to flow through the ``side" pipe. The similarity is in that the flow in the side pipe is then independent of the direction in which the fluid flows in the main pipe. Note that the rate of the ``sucked-in" flow is controlled by the hydrodynamic velocity at the junction, which is sensitive to a local expansion or narrowing of the flow (``Venturi effect"), but the phenomenon itself is generic, in similarity to the generic nature of emergent chirality in the interacting electron system.

\acknowledgments

P.W. acknowledges discussions with F.~Evers as well as support through Distinguished Senior Fellowship at Karlsruhe Institute of Technology. This work was supported by the Russian Foundation for Basic Research under Grant No.\ 18-02-01016.


\begin{thebibliography}{99}

\bibitem{Kane1992} C.L. Kane and M.P.A. Fisher, Phys.\ Rev.\ B \textbf{46}, 15233 (1992).

\bibitem{Furusaki1993} A. Furusaki and N. Nagaosa, Phys.\ Rev.\ B {\bf 47}, 4631 (1993).

\bibitem{Yue1994} D. Yue, L.I. Glazman, and K.A. Matveev, Phys.\ Rev.\ B {\bf 49}, 1966 (1994).

\bibitem{nayak99} C. Nayak, M.P.A. Fisher, A.W.W. Ludwig, and H.H. Lin, Phys.\ Rev.\ B {\bf 59}, 15694 (1999).

\bibitem{Lal2002} S. Lal, S. Rao, and D. Sen, Phys.\ Rev.\ B {\bf 66}, 165327 (2002).

\bibitem{chen02} S. Chen, B. Trauzettel, and R. Egger, Phys.\ Rev.\ Lett.\ {\bf 89}, 226404 (2002).

\bibitem{Oshikawa2006} M. Oshikawa, C. Chamon, and I. Affleck, J. Stat.\ Mech.\ (2006) P02008.

\bibitem{Aristov2009} D.N. Aristov and P. W\"olfle, Phys. Rev. B {\bf 80}, 045109 (2009).

\bibitem{Aristov2010} D.N.~Aristov, A.P.~Dmitriev, I.V.~Gornyi, V.Y. Kachorovskii, D.G.~Polyakov, and P.~W\"olfle, Phys.\ Rev.\ Lett.\ {\bf 105}, 266404 (2010).

\bibitem{hou12} C.-Y. Hou, A. Rahmani, A.E. Feiguin, and C. Chamon, Phys.\ Rev.\ B {\bf 86}, 075451 (2012).

\bibitem{Tomonaga1950} S. Tomonaga, Prog.\ Theor.\ Phys.\ {\bf 5}, 544 (1950).

\bibitem{Luttinger1963} J.M. Luttinger, J. Math.\ Phys.\ {\bf 4}, 1154 (1963).

\bibitem{Giamarchi2004} T. Giamarchi, \textit{Quantum Physics in One Dimension} (Oxford University Press, Oxford, 2004).

\bibitem{Aristov2013} D.N. Aristov and P. W\"olfle, Phys.\ Rev.\ B {\bf 88}, 075131 (2013).

\bibitem{Sassetti1996} M. Sassetti and B. Kramer, Phys.\ Rev.\ B {\bf 54}, R5203 (1996).

\bibitem{Furusaki1996} A. Furusaki and N. Nagaosa, Phys.\ Rev.\ B {\bf 54}, R5239 (1996).

\bibitem{Egger2000} R. Egger, H. Grabert, A. Koutouza, H. Saleur, and F. Siano, Phys.\ Rev.\ Lett.\ {\bf 84}, 3682 (2000).

\bibitem{Dolcini2003} F. Dolcini, H. Grabert, I. Safi, and B. Trauzettel, Phys.\ Rev.\ Lett.\ {\bf 91}, 266402 (2003).

\bibitem{Metzner2012} W. Metzner, M. Salmhofer, C. Honerkamp, V. Meden, and K. Sch\"onhammer, Rev.\ Mod.\ Phys.\ {\bf 84}, 299 (2012).

\bibitem{Aristov2017} D.N. Aristov, I.V. Gornyi, D.G. Polyakov, and P. W\"olfle, Phys.\ Rev.\ B {\bf 95}, 155447 (2017).

\bibitem{Aristov2018} D.N. Aristov and P. W\"olfle, Phys. Rev. B {\bf 97}, 205101 (2018).

\bibitem{Safi1995} I. Safi and H.J. Schulz, Phys.\ Rev.\ B {\bf 52}, R17040 (1995).

\bibitem{maslov95} D.M. Maslov and M. Stone, Phys.\ Rev.\ B {\bf 52}, R5539 (1995).

\bibitem{ponomarenko95} V.V. Ponomarenko, Phys.\ Rev.\ B {\bf 52}, R8666 (1995).

\bibitem{egger96} R. Egger and H. Grabert, Phys.\ Rev.\ Lett.\ {\bf 77}, 538 (1996); Phys.\ Rev.\ B {\bf 58}, 10761 (1998).

\bibitem{alekseev96} A.Yu. Alekseev, V.V. Cheianov, and J. Fr\"ohlich, Phys.\ Rev.\ B {\bf 54}, R17320 (1996).

\bibitem{chamon97} C. de C. Chamon and E. Fradkin, Phys.\ Rev.\ B {\bf 56}, 2012 (1997).

\bibitem{gutman10} The bosonization approach becomes much more involved for the case of a non-Fermi-Dirac distribution of electrons supplied by the reservoir, see D.B. Gutman, Y. Gefen, and A.D. Mirlin, Phys.\ Rev.\ B {\bf 81}, 085436 (2010).

\bibitem{Aristov2014} D.N. Aristov and P. W\"olfle, Phys.\ Rev.\ B {\bf 90}, 245414 (2014).

\bibitem{Shi2016} Z. Shi and I. Affleck, Phys.\ Rev.\ B {\bf 94}, 035106 (2016); Z. Shi, J. Stat.\ Mech.\ (2016) 063106.

\bibitem{remark3} To avoid terminological confusion, notice that the notion of chirality in the term ``chiral fixed point" in Ref.~\cite{Oshikawa2006} refers to the {\it circular} current flowing (also at equilibrium) around the ``hole" of a not simply connected triple junction in the absence of time-reversal symmetry. This should not be confused with the notion of chirality in the term ``chiral current" in the lead, which we introduce in this paper with regard to the persistent ``unidirectionality" of the current, as illustrated in Fig.~\ref{f1}.

\bibitem{aristov11} D.N. Aristov and P. W\"olfle, Phys.\ Rev.\ B {\bf 84}, 155426 (2011).

\bibitem{belinicher80} V.I. Belinicher and B.I. Sturman, Sov.\ Phys.\ Usp.\ {\bf 23}, 199 (1980).

\bibitem{zyuzin04} B. Spivak and A. Zyuzin, Phys.\ Rev.\ Lett.\ {\bf 93}, 226801 (2004); E. Deyo, B. Spivak, and A. Zyuzin, Phys.\ Rev.\ B {\bf 74}, 104205 (2006) proposed a mechanism for the ``magnetochiral" effect (the current is an odd function of the magnetic field and an even function of the voltage) based on the combined action of electron-electron interaction and nonequilibrium. However, despite the similarity in this respect, this magnetochiral effect, which combines features of both the photogalvanic and Hall effects, necessarily relies on the absence of both inversion symmetry (in a mesoscopic disordered sample for a given realization of disorder) and time-reversal symmetry at the level of the Hamiltonian, in essential contrast to emergent chirality proposed in our paper.

\bibitem{feldman} Rectification effects in junctions of TLL wires were previously considered within the bosonization framework, e.g., in Refs.~\cite{feldman05} and \cite{wang11}. The rectification mechanisms discussed in Ref.~\cite{wang11} for the three-lead case are not, however, related to emergent chirality.

\bibitem{feldman05} D.E. Feldman, S. Scheidl, and V.M. Vinokur, Phys.\ Rev.\ Lett.\ {\bf 94}, 186809 (2005).

\bibitem{wang11} C. Wang and D.E. Feldman, Phys.\ Rev.\ B {\bf 83}, 045302 (2011).

\bibitem{remark1} For the complex behavior of the flow trajectory around point $A$, in particular, the sharp change of the trajectory between attraction and the runaway, see the discussion below Eq.~(7) in Ref.~\cite{Aristov2010}. Note also that the power-law exponent $\alpha_3$ for $\bar{G}_b$ in Eq.~(\ref{89}) assumes that $|\alpha_3|\gg\alpha^2$ \cite{Aristov2010,aristov11}.

\bibitem{buccheri18} A similar construction for spin currents in a not simply connected triple junction of spin chains with time-reversal symmetry broken at the junction was introduced in F. Buccheri, R. Egger, R.G. Pereira, and F.B. Ramos, Phys.\ Rev.\ B {\bf 97}, 220402 (2018); Nucl.\ Phys.\ B {\bf 941}, 794 (2019).

\bibitem{remark2} It might be interesting to note that the effective magnetic flux in Eq.~(\ref{95}) corresponds to a large effective magnetic field. For example, for $\alpha_3=0.3$ and the junction area $10\times 10\,{\rm nm}^2$, the magnetic field is about 20\,T.


\end{thebibliography}
\end{document}